\documentclass[aps,prb,twocolumn,superscriptaddress]{revtex4-1} 
\usepackage[T1]{fontenc}
\usepackage{amssymb}

\usepackage{hyperref} 
\hypersetup{
    bookmarks=true,         
    unicode=true,          
    pdftoolbar=false,        
    pdfmenubar=true,        
    pdffitwindow=true,      
    pdftitle={My title},    
    pdfauthor={Author},     
    pdfsubject={Subject},   
    pdfnewwindow=true,      
    pdfkeywords={keywords}, 
    colorlinks=true,       
    linkcolor=red,          
    citecolor=black,        
    filecolor=magenta,      
    urlcolor=cyan           
}

\usepackage[all]{hypcap}  
\usepackage{graphicx}
\usepackage{amsmath}
\usepackage{natbib}
\usepackage{layouts}
\usepackage{verbatim}

\usepackage{mathtools}

\DeclarePairedDelimiter\floor{\lfloor}{\rfloor}
 
\begin{document}

\title{Quantum oscillations in the kinetic energy density:\\Gradient corrections from the Airy gas} 

\author{A.~Lindmaa} 
\email{alexander.lindmaa@liu.se}
\affiliation{Department of Physics, Chemistry and Biology (IFM), Link\"oping University, SE-581 83  Link\"oping, Sweden}
\author{A.~E.~Mattsson}
\email{aematts@sandia.gov}
\affiliation{Multi-Scale Science MS 1322, Sandia National Laboratories, Albuquerque, New Mexico, 87185-1322, USA}
\author{R.~Armiento}
\email{rickard.armiento@liu.se}
\affiliation{Department of Physics, Chemistry and Biology (IFM), Link\"oping University, SE-581 83 Link\"oping, Sweden}
\date{\today}

\begin{abstract} 

We derive a closed form expression for the quantum corrections to the kinetic energy density (KED) in the Thomas-Fermi (TF) limit of a linear potential model system in three dimensions (the Airy gas). The universality of the expression is tested numerically in a number of three dimensional model systems: (i) jellium surfaces, (ii) hydrogen-like potentials, (iii) systems confined by an harmonic potential in one and (iv) all three dimensions, and (v) a system with a cosine potential (the Mathieu gas). Our results confirm that the usual gradient expansion of extended Thomas-Fermi theory (ETF) does not describe the quantum oscillations for systems that incorporate surface regions where the electron density drops off to zero. We find that the correction derived from the Airy gas is universally applicable to relevant spatial regions of systems of type (i), (ii), and (iv), but somewhat surprisingly not (iii). We discuss possible implications of our findings to the development of functionals for the kinetic energy density. 
\end{abstract} 

\maketitle  

\section{Introduction} 
 
Since the early days of quantum mechanics there has been an interest in accurately describing the kinetic energy (KE) of a system of non-interacting fermions given the particle density. Such descriptions have been paramount in the development of schemes for computations of physical properties of atoms, molecules and solids which are in ubiquitous use today across disciplines. The derivation and evaluation of approximate expressions of the kinetic energy is still an active area of research with applications in, e.g., orbital-free (OF) density functional theory (DFT) \cite{hohenberg_inhomogeneous_1964, kohn_self-consistent_1965}, high-temperature applications of DFT, and as an intermediate step in developing improved approximations for the exchange-correlation energy. Applications are also found in the field of nuclear DFT \cite{nuclear_dft} and trapped degenerate fermion gases\cite{PhysRevLett.109.030401}. 

A common starting point for most approximations of the KE is Thomas-Fermi (TF) theory \cite{thomas_calculation_1927, fermi_metodo_1927}, which is exact for a uniform electron gas. A number of historically important works have derived corrections to TF for a weakly inhomogeneous electron system as an expansion in gradients of the electron density \cite{kirzhnits, hodges1973, brack_extended_1976, hohenberg_inhomogeneous_1964, yang_gradient_1986}, and we will refer to this as the extended TF (ETF) gradient expansion (GE). However, it has also been noted that for systems with surface regions, i.e., regions where the electron density drops to zero, the ETF GE of the kinetic energy density is not valid and further corrections of the same order in the density and density gradient are necessary (see, e.g., Refs.~\onlinecite{kohn_quantum_1965, elliott_semiclassical_2008} and references therein for an extended discussion.) Despite a frequent appeal to TF and ETF theory in the literature, the need for such corrections is rarely discussed. 

In the present paper we utilize a closed form expression of the non-interacting KE of the Airy gas (AG) surface model system \cite{kohn_edge_1998} to derive a modified ETF GE that includes the quantum corrections. When applying the obtained expression to other model systems with electron surfaces, e.g., the jellium surface model, hydrogen-like potentials, and a system confined by an isotropic harmonic potential, we find these quantum corrections to provide a crucial correction to the local description of the kinetic energy density. However, for two other systems that we investigate, i.e., a system with a cosine potential, and a system confined by an harmonic potential in only one dimension, we find that neither of the two expansions holds unreservedly.

The rest of the paper is organized as follows: in Sec.~II we summarize the important equations for the kinetic energy, the edge electron gas class of model systems and the Airy gas model. In Sec.~III we derive the central expression of the work, a gradient expansion of the kinetic energy density which includes the quantum corrections from the Airy gas surface. In Sec.~IV we investigate the universality of the obtained expressions in a range of numerical tests. In Sec.~V we discuss our findings. Finally, Sec.~VI presents a summary and the main conclusions of this work.

\section{Background} 

Our primary interest for the KE of non-interacting fermions is due to its central importance in Kohn-Sham (KS) DFT \cite{kohn_self-consistent_1965}, and thus we will adopt the relevant terminology from this field. Hence, consider a system of $N$ non-interacting fermions with ground state particle density $n(\mathbf{r})$ which we assume to be continuous everywhere. The KE is given exactly in Hartree atomic units by the functional 
\begin{equation}\label{eq:Ts}
T_{\mathrm{s}}[n]= -\frac{1}{2} \sum_{\nu: \epsilon_{\nu} \leq \mu} \int \psi_{\nu}^{*}(\boldsymbol{r}) \nabla^2 \psi_{\nu}(\boldsymbol{r}) d^3\boldsymbol{r},
\end{equation} 
where $\nu$ indexes available states including the spin degree of freedom, $\mu$ is the self-consistent chemical potential and $\{\psi_{\nu}\}_{\nu=1}^{\infty}$ are the KS orbitals with corresponding eigenvalues $\{\epsilon_{\nu}\}_{\nu=1}^{\infty}$. The KS-orbitals are formally functionals of $n(\boldsymbol{r})$, making Eq.~(\ref{eq:Ts}) an implicit functional. Equation (\ref{eq:Ts}) directly defines one possible kinetic energy density (KED) as 
\begin{equation}
\tau_1(\boldsymbol{r})=-\frac{1}{2} \sum_{\nu: \epsilon_{\nu} \leq \mu} \psi_{\nu}^*(\boldsymbol{r}) \nabla^2 \psi_{\nu}(\boldsymbol{r}).
\end{equation} 
The KED is only unique up to a term that integrates to zero over the system in Eq.~(\ref{eq:Ts}). In a closed or periodic system, Gauss' divergence theorem applied to $\nabla n$ gives that the Laplacian of the electron density $\nabla^2 n$ integrates to zero. This can be exploited to construct the following alternative KED which is positive at all points in space
\begin{equation}
\tau(\boldsymbol{r}) = \tau_1(\boldsymbol{r}) + \frac{1}{4}\nabla^2n(\boldsymbol{r}) = \frac{1}{2}\sum_{\nu: \epsilon_{\nu} \leq \mu} |\nabla \psi_{\nu}(\boldsymbol{r})|^2. \label{eq:pos_ked}
\end{equation} 
This choice of KED is advantageous for developing approximations, since an approximation that fulfills the constraint $\tau(\boldsymbol{r}) \geq 0$ will avoid the unphysical result $T_{\mathrm{s}} < 0$ for all $n(\boldsymbol{r})$. We stress that Eq.~(\ref{eq:pos_ked}) is taken as a definition, which makes $\tau$ valid without ambiguity even for open non-periodic systems. However, the caveat for such systems is that $\tau$ is then not strictly a KED, since integration will not give $T_{\mathrm{s}}$ (one would instead need to integrate $\tau(\boldsymbol{r}) - (1/4)\nabla^2n(\boldsymbol{r})$.) 

A starting point for many approximations of $\tau$ is the TF approximation,
\begin{equation} 
\tau^{\mathrm{TF}}[n] = C_{\mathrm{TF}} n^{\frac{5}{3}}(\boldsymbol{r}),
\end{equation}
with $C_{\mathrm{TF}} = (3/10)(3\pi^2)^{\frac{2}{3}}$. Based on the scaling relation of the total non-interacting kinetic energy \cite{engel_dreizler} one can assume $\tau$ to be homogeneous of degree $5/3$ under uniform coordinate scaling \cite{parr_yang}. Hence, a (semi-)local density-functional approximation (DFA) of the KED takes the form
\begin{equation} 
\tau^{\mathrm{DFA}}[n] = \tau^{\mathrm{TF}}[n] F^{\mathrm{DFA}}_{\tau}(s,q, \ldots), 
\end{equation}
where $F^{\mathrm{DFA}}_\tau$ is the refinement factor, and the scaled gradient $s$ and Laplacian $q$ are 
\begin{eqnarray}
s= \frac{|\nabla n(\boldsymbol{r})|}{2(3\pi^2)^{\frac{1}{3}}n^{\frac{4}{3}}(\boldsymbol{r})}, \\
q= \frac{\nabla^2 n(\boldsymbol{r})}{4(3\pi^2)^{\frac{2}{3}}n^{\frac{5}{3}}(\boldsymbol{r})}.
\end{eqnarray} 
We define as a \emph{limit of slowly varying density} any limit where $s, q$ and all higher order terms $\to 0$.

Early efforts to find improved approximations of $\tau$ date back to 1935, when von Weizs\"acker\cite{weizsacker_1935} derived a KED approximation with $F_\tau^{\textrm{vW}} = (5/3)\, s^2$. In 1957 Kirzhnits \cite{kirzhnits, hodges1973} used commutator operator formalism to derive gradient corrections to TF theory in the limit of a weakly perturbed uniform electron gas. A number of extensions to the original result have followed. Hohenberg and Kohn\cite{hohenberg_inhomogeneous_1964} developed a density gradient technique based on linear response formalism. An expansion in $\hbar$ of the Green's function representation due to Wigner and Kirkwood has also been calculated\cite{jennings_bhaduri}. In a paper by Yang, corrections are derived in terms of the Green's function from the first-order reduced density matrix\cite{yang_gradient_1986}. From these past works it is well established that the ETF gradient correction to TF, i.e., the gradient corrections for the slowly varying limit of a weakly perturbed uniform electron gas, are to second order
\begin{equation}\label{eq:etf_grad} 
F_\tau^{\mathrm{ETF}}(s,q) = 1 + \frac{5}{27}s^2 + \frac{20}{9}q.
\end{equation} 

However, as explained in the introduction, it has been observed that the ETF expansion does not apply universally to regions of slowly varying electron density in a system that is not a weakly perturbed uniform electron gas, but rather has surface regions, i.e., regions where the electron density drops to zero (see, e.g., Refs.~\onlinecite{kohn_quantum_1965, elliott_semiclassical_2008}). The main idea put forward in this paper is to derive explicit corrections from a surface model system (the AG) and investigate how general the resulting expression is when applied to other model systems. 

We note that the kinetic energy of the AG model and related systems have been discussed in other recent works. Vitos \emph{et al.} \cite{PhysRevA.61.052511} and Constantin and Ruzsinszky\cite{PhysRevB.79.115117} have both presented KED functionals based on parameterizations of the AG KED. Both works discuss the role of the Laplacian term in achieving an optimal local description of the KED across the surface. Constantin and Ruzsinszky specifically enforce the ETF GE in the limit of slowly varying density. In contrast, in the present work our focus is the exact behavior in the limit of slowly varying electron density far inside the surface. 

\subsection{Electronic edges and the AG} 

The derivations in the present work start from the formalism and results by Kohn and Mattsson  \cite{kohn_edge_1998}, which are outlined in the following. We start from the general model system of an edge electron gas (EG) taken to be an inhomogeneous system of electrons with an \emph{electronic edge}. The edge is defined in terms of the classical turning points
\begin{equation}
v_{\mathrm{s}}({\boldsymbol{r}})=\mu, 
\end{equation}
where $v_{\mathrm{s}}$ is the KS effective potential and $\mu$ is between the highest occupied and lowest unoccupied KS eigenvalue. This describes a surface (in the mathematical sense) outside of which the orbitals decay at exponential rate. We take $v_{\mathrm{s}}$ to be constant in two spatial directions whilst varying in the third. The resulting KS-orbitals are labeled with the quantum numbers $\nu = (k_1, k_2, \eta)$, where the $k_i$ are plane wave numbers for the $x$- and $y$-dimensions and $\eta$ is the quantum number associated with the $z$-dimension. Note that $\eta$ can either be a continuous or discrete quantum number, depending on the energy spectrum of the potential across the surface. In Ref.~\onlinecite{kohn_edge_1998} the electron density of the EG for a general $v_{\mathrm{s}}(z)$ is found to be 
\begin{equation}\label{eq:eg_density}
n^{\mathrm{EG}}(\boldsymbol{r}) = \frac{1}{\pi} \sum_{\eta: \epsilon_{\eta} \leq \mu}\varphi^2_{\eta}(z)|\mu - \epsilon_{\eta}|,  
\end{equation}
where $\varphi_{\eta}$ are the corresponding eigenfunctions in the $z$-direction.  

Furthermore, the AG is an edge gas with a linear potential 
\begin{eqnarray}   
v^{\mathrm{AG}}_{\mathrm{s}}(\boldsymbol{r}) = \bigg \{ \begin{array}{rl}
 Fz \, &\mbox{  $ z \ge -L $} \\ 
  +\infty \, &\mbox{\,  \text{otherwise}}
       \end{array},
\end{eqnarray} 
where $F>0$ is the slope in the potential. This slope defines a characteristic length $l \equiv \sqrt[3]{\frac{1}{2F}}$. Apart from Ref.~\onlinecite{kohn_edge_1998}, similar models have been investigated also in a number of previous works\cite{baltin_1971, baltin_1972, PhysRevB.79.115117, PhysRevLett.68.1359, sahni_1997, PhysRevA.61.052511}. The KS equation 
\begin{equation} 
\left(-\frac{1}{2} \frac{d^2}{dz^2}+Fz\right)\varphi_{\eta}=\epsilon_{\eta}\varphi_{\eta}(z), 
\end{equation}
with $\varphi_{\eta}(-L) = \varphi_{\eta}(\infty)=0$ has the solutions
\begin{equation}\label{eq:ag_orbitals} 
\varphi_{\eta}(z)=\frac{\sqrt{\pi}}{\sqrt[4]{Ll}}\text{Ai}\left(\frac{z}{l}+\frac{\epsilon_{\eta}}{\epsilon}\right), 
\end{equation}
and eigenvalues 
\begin{equation}
\epsilon_{\eta}=-\eta \sqrt{\frac{l}{L}} \pi \tilde \epsilon,\ \textrm{with}\ \tilde \epsilon \equiv \sqrt[3]{\frac{F^2}{2}},
\end{equation}
where $\mathrm{Ai}$ and $\mathrm{Bi}$ are the Airy functions. The orbitals and eigenvalues scale directly with $l$, which makes the AG effectively a zero parameter model. The absolute energy scale is chosen to make the chemical potential equal to zero. The eigenvalues $\{\epsilon_{\eta}\}_{\eta=1}^{\infty}$ are equally spaced and form a countable set. We introduce the scaled coordinate and scaled eigenvalues as, respectively,
\begin{equation} 
\zeta \equiv z/l, \quad \epsilon \equiv\epsilon_{\eta}/\tilde \epsilon. 
\end{equation}
As we take the limit $L \rightarrow \infty$ the eigenvalues turn into a continuous eigenspectrum, i.e., 
\begin{equation}\label{eq:sum_integral} 
 \frac{\Delta \epsilon_{\eta}}{\tilde{\epsilon}} \rightarrow d\epsilon \implies \sum_{\epsilon_{\eta}} \frac{\Delta \epsilon_{\eta}}{\tilde{\epsilon}}  \rightarrow \int d\epsilon.
\end{equation}
Using Eqs.~(\ref{eq:ag_orbitals}) and (\ref{eq:sum_integral}) in Eq.~(\ref{eq:eg_density}) gives the AG electron density,
\begin{equation}
n_0^{\mathrm{AG}}(\zeta)=\frac{1}{2\pi} \int^{\infty}_{0} d\epsilon \, \epsilon \, \text{Ai}^2(\zeta - \epsilon), 
\end{equation} 
with $n^{\mathrm{AG}}(z/l) = ( 1 / l^3 )n^{\mathrm{AG}}_0(\zeta)$. By Eq.~(A.4) in Ref.~\onlinecite{albright_1977} we then have
\begin{equation}
n^{\mathrm{AG}}_0(\zeta) = \frac{1}{6\pi}\bigg[ 2\zeta^2 \text{Ai}^2(\zeta) - \text{Ai}(\zeta)\text{Ai}^{\prime}(\zeta) - 2\zeta \text{Ai}^{\prime2}(\zeta) \bigg].
\end{equation}
Successive differentiation with respect to $\zeta$ gives for the dimensionless scaled gradient and Laplacian respectively 
\begin{eqnarray} 
s^{\mathrm{AG}}(\zeta) = \frac{1}{2\pi} \frac{\text{Ai}^{\prime 2}(\zeta) - \zeta \text{Ai}^2(\zeta)}{2(3\pi^2)^{\frac{1}{3}}[n_0^\mathrm{AG}(\zeta)]^{\frac{4}{3}}}, \\
q^{\mathrm{AG}}(\zeta) = \frac{1}{2\pi} \frac{\text{Ai}^2(\zeta)}{4(3\pi^2)^{\frac{2}{3}}[n_0^{\mathrm{AG}}(\zeta)]^{\frac{5}{3}}}.
\end{eqnarray} 
where we note that $\zeta \to -\infty \Rightarrow s, q \to 0$, which means that the far inner region of the AG is a limit of slowly varying density. However, this limit of slowly varying density is fundamentally different from that in a weakly perturbed uniform electron gas. 

\section{The KED in the Airy gas} 

In this section we derive the central result of this work, a GE based on the AG model system, i.e., a GE up to second order in $s$ and $q$ that includes the quantum corrections due to the surface. Starting from the definition of the positive KED in Eq.~(\ref{eq:pos_ked}) and using Eq.~(\ref{eq:eg_density}) gives
\begin{equation}\label{eq:ked_eg}
\tau^{\mathrm{EG}}(z)=\frac{1}{2}\sum_{\eta: \epsilon_{\eta} \leq \mu }\left[ \frac{|\mu - \epsilon_{\eta}|^2}{2\pi}\varphi_{\eta}^2(z) + \frac{|\mu - \epsilon_{\eta}|}{2\pi} \varphi^{\prime 2}_{\eta}(z) \right], 
\end{equation} 
where we have used that the orbital energies satisfy $\frac{1}{2} \left( k_1^2 + k_2^2 \right) + \epsilon_{\eta} \leq \mu$. Equation~(\ref{eq:ked_eg}) is a general expression for the positive KED of an EG. As was previously mentioned, we take for the AG $\mu = 0$.   

Inserting the AG orbitals of Eq.~(\ref{eq:ag_orbitals}) into Eq.~(\ref{eq:ked_eg}) and taking $L \rightarrow \infty$ results in an expression for the KED in terms of integrals over Ai functions\cite{PhysRevA.61.052511, PhysRevB.79.115117} 
 \begin{eqnarray} 
\tau^{\mathrm{AG}}_0(\zeta) &=& \bigg[\frac{1}{8\pi} \int_{0}^{\infty} d\epsilon \, \epsilon^2 \text{Ai}^2(\zeta-\epsilon) \nonumber \\ 
&+& 
\frac{1}{4\pi} \int_{0}^{\infty} d\epsilon \, \epsilon \bigg(\frac{d}{d\zeta}\text{Ai}(\zeta-\epsilon)\bigg)^2\bigg],
\end{eqnarray}
where $\tau^{\mathrm{AG}}(z/l)=l^{-5}\tau_0^{\mathrm{AG}}(\zeta)$. Using Eqs.~(A.6) and (A.7) found in Ref.~\onlinecite{albright_1977}, we arrive at
\begin{eqnarray}
\tau^{\mathrm{AG}}_{0}(\zeta) &=& \frac{1}{20\pi}\bigg[2(1-\zeta^3)\text{Ai}^2(\zeta) + \zeta \text{Ai}(\zeta) \text{Ai}^{\prime}(\zeta) \nonumber \\ 
&+& 
2\zeta^2\text{Ai}^{\prime}(\zeta)\bigg] . 
\end{eqnarray}
This is an exact expression on closed form valid throughout the AG system.
We are interested in the quantum oscillations in the regime of slowly varying electron density, i.e., where $s, q \to 0$ far inside the classically allowed region of the system.    

In this limit, both $n^{\mathrm{AG}}_0(\zeta)$ and $\tau^{\mathrm{AG}}_0(\zeta)$ are \emph{unbounded} continuous functions. Moreover, $\tau^{\mathrm{TF}}[n_{0}^{\mathrm{AG}}]$ is continuous and the exact refinement factor
\begin{equation}\label{eq:Fagked}
F^{\mathrm{AG}}(\zeta) = \frac{\tau^{\mathrm{AG}}_{0}(\zeta)}{\tau^{\mathrm{TF}}[n_{0}^{\mathrm{AG}}]} 
\end{equation}
is bounded and analytic as $\zeta \rightarrow -\infty$. Expanding Eq.~(\ref{eq:Fagked}) in a series around $\zeta = -\infty$ gives 
\begin{eqnarray}\label{eq:ref_exp}
F^{\mathrm{AG}}(\zeta) &=& 1 + \frac{5 \left(5 + 6\sin \left( \frac{4}{3}\zeta^{\frac{3}{2}} \right) \right)}{48 \zeta^3} \nonumber \\ 
&-& \frac{55}{192}\left( \frac{1}{\zeta} \right)^{\frac{9}{2}}\cos \left(\frac{4}{3}\zeta^{\frac{3}{2}} \right) + \mathcal{O}\left( \frac{1}{\zeta^5} \right).
\end{eqnarray}
This expansion has previously been discussed by Baltin\cite{baltin_1972}. We now proceed by expanding the scaled gradient $s^{\mathrm{AG}}(\zeta)$ and Laplacian $q^{\mathrm{AG}}(\zeta)$ in a Taylor series in the same way. This gives us 
\begin{equation}\label{eq:s_exp} 
s^{\mathrm{AG}}(\zeta) = \frac{3}{4} \left (\frac{1}{\zeta} \right)^{\frac{1}{3}} - \frac{3\cos{\left(\frac{4}{3}\zeta^{\frac{3}{2}}\right)}}{16\zeta^3} + \mathcal{O}\left( \frac{1}{\zeta^{\frac{1}{9}}} \right) 
\end{equation}
and
\begin{eqnarray}\label{eq:q_exp} 
q^{\mathrm{AG}}(\zeta) &=& \frac{3 \left( 1 + \sin \left( \frac{4}{3} \zeta^{\frac{3}{2}} \right) \right)}{16 \zeta^3} \nonumber \\ 
&-& \frac{5}{128}  \left( \frac{1}{\zeta} \right)^{\frac{9}{2}} \cos \left( \frac{4}{3} \zeta^{\frac{3}{2}} \right) + \mathcal{O}\left( \frac{1}{\zeta^6} \right) 
\end{eqnarray} 
respectively. We note that Eqs.~(\ref{eq:ref_exp}) and (\ref{eq:q_exp}) to leading order both contain terms proportional to
\begin{equation}
\frac{1}{\zeta^3}\sin{\left( \frac{4}{3}\zeta^{\frac{3}{2}} \right)}.
\end{equation}
Hence, by identification we can extract the coefficient for a term proportional to the scaled Laplacian $q_{\mathrm{AG}}(\zeta)$. We get to leading order
\begin{equation}\label{eq:bare_term} 
F^{\mathrm{AG}}(\zeta) = 1 + \frac{10}{3}q^{\mathrm{AG}}(\zeta) - \frac{}{}\frac{5}{48}\frac{1}{\zeta^3}.
\end{equation}
The leading term in Eq.~(\ref{eq:s_exp}) is non-oscialltory, and the expression can thus be inverted to give to leading order in $s$
\begin{equation}\label{eq:inverse} 
\zeta^{\mathrm{AG}}(s) = \frac{1}{2} \left( \frac{9}{2s^2} \right)^{\frac{1}{3}}
\end{equation}
If we substitute Eq.~(\ref{eq:inverse}) into Eq.~(\ref{eq:bare_term}) we finally obtain, to second order in $|\nabla n|$, the expression
\begin{equation}\label{eq:ag_corrections} 
F(s,q) =1 - \frac{}{}\frac{5}{27}s^2 + \frac{10}{3}q  
\end{equation} 
The derivation of Eq.~(\ref{eq:ag_corrections}) is based on the identification of terms between Eqs.~(\ref{eq:ref_exp})--(\ref{eq:q_exp}), but is unique in the sense that no other expression with only linear dependence on $s^2$ and $q$ (i.e., to second order in $|\nabla n|$) can reproduce exactly the oscillatory first-order term in Eq.~(\ref{eq:ref_exp}). The result is a refinement function derived from the limit of slowly varying density far inside the surface from a linear potential, i.e., it is a GE that includes quantum corrections from a surface. Note that the term proportional to $s^2$ in Eq.~(\ref{eq:ag_corrections}) has the same coefficient as the ETF GE given by Eq.~(\ref{eq:etf_grad}), but differs in sign, whereas the term proportional to $q$ is completely different. The expression is a main result of this paper, and we will refer to it as the AG-GE. We are not aware of prior works that address the system dependence of GEs with quantum corrections, so there is no \emph{a priori} reason to expect this expression to be broadly applicable to a large set of systems with surfaces. Nevertheless, in the following we will investigate the possible generality of this expression in regions of slowly varying electron density in other model systems. However, our first numerical investigation is the behavior of the ETF GE and the AG-GE for the actual AG model system itself.

\begin{figure*} 
  \begin{tabular}{c @{\hspace{0.05\linewidth}} c}
    \multicolumn{1}{l}{a)} & \multicolumn{1}{l}{b)} \\
    \includegraphics[]{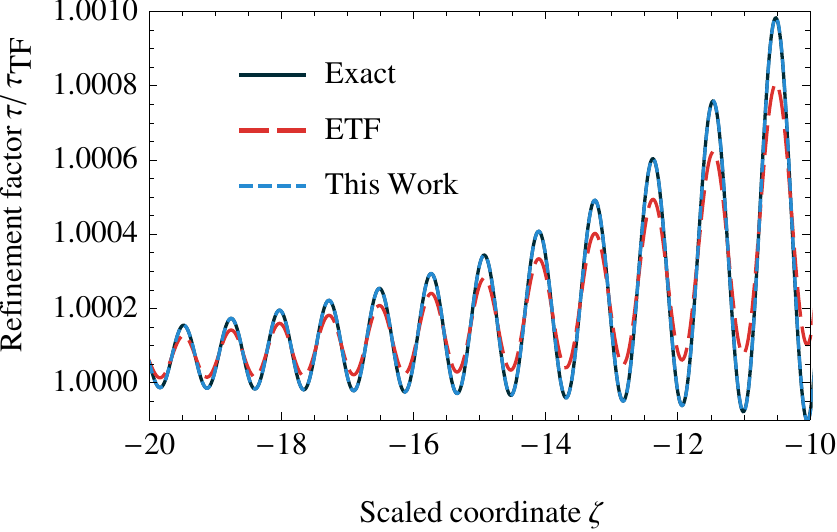} &
    \includegraphics[]{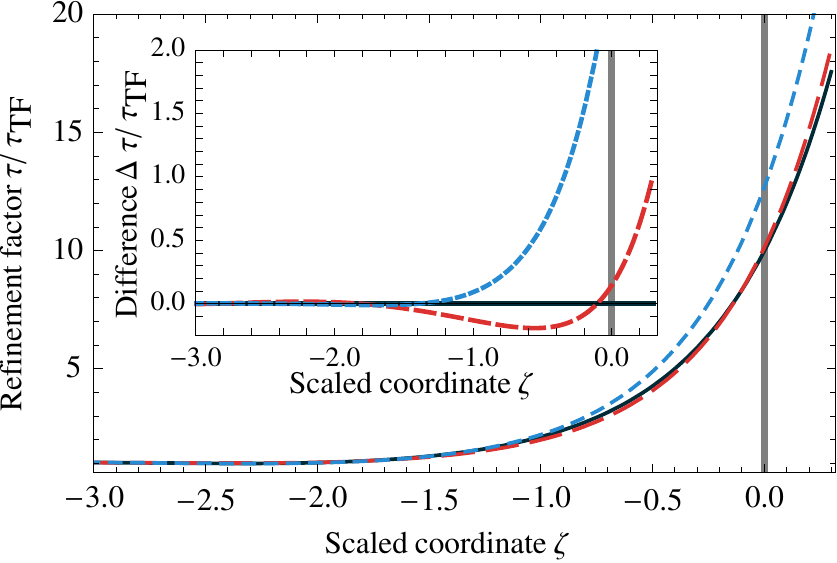} \\
   \multicolumn{1}{l}{c)} & \multicolumn{1}{l}{d)} \\
    \includegraphics[]{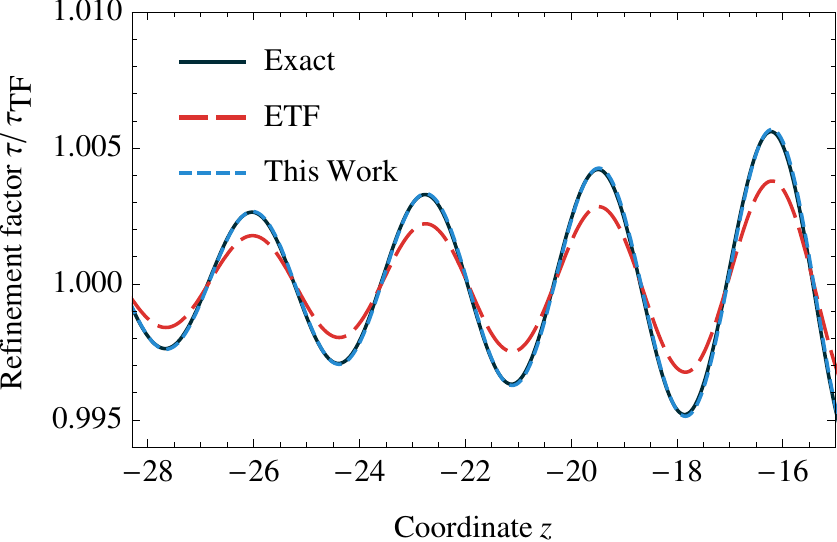} &
    \includegraphics[]{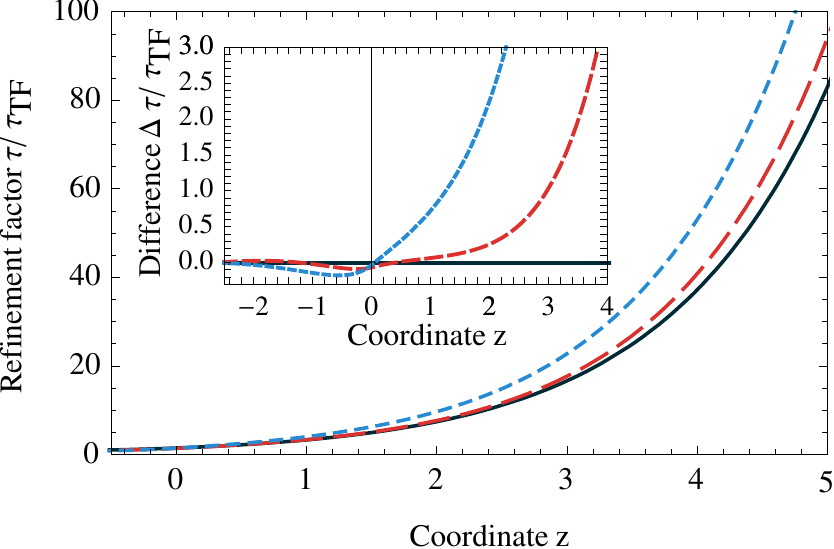} \\
  \end{tabular}
  \caption{The ETF GE and the AG-GE in Eq.~(\ref{eq:ag_corrections}) compared to the exact KED for the AG (a,b) and for a jellium surface with $r_s=1$ (c,d). The AG system shown in the two topmost panels (a,b) is the system used to derive the AG-GE. These panels are placed side-by-side with the corresponding application of the expressions in the jellium system (c,d) for comparison. (a) The behavior far inside the AG surface ($\zeta \to -\infty$) as a function of the scaled coordinate $\zeta$. The GE derived from the AG unsurprisingly describes the oscillations in the KED well, whereas the ETF GE does not. (b) The behavior in the surface region of the AG. The vertical line shows the position of the classical turning point. The inset shows the difference between the two expansions and the exact KED. (c) The regime of slowly varying electron density at large negative $z$ of the jellium model. Also for the jellium system the AG-GE describes the oscillations in the KED well, whereas the ETF GE fails to do so. This mimics closely the behavior seen for the AG in panel (a). (d) The surface region of the jellium model. In both (b) and (d) the ETF GE deviates less than the AG-GE from the exact KED over the surface region.}  
  \label{fig:fig1}
\end{figure*}

In Fig.~\ref{fig:fig1} (a) and (b) we compare numerically the exact AG KED, the AG-GE, and the ETF GE far inside the inner region of the AG, and across the surface region. We see that the correction provided by Eq.~(\ref{eq:ag_corrections}) over ETF is crucial to properly account for the oscillations in the KED far from the surface. In the edge region both of the expansions are expected to fail, since in this region the density is not slowly varying. However, it appears the quantum corrections included in Eq.~(\ref{eq:ag_corrections}) worsen the result in this region compared to ETF. 

\section{Numerical Results} 

In the following subsections we will explore the validity of Eq.~(\ref{eq:ag_corrections}) in comparison with the ETF GE for a number of model systems. 

\subsection{The jellium surface model}

We turn first to the jellium surface model system \cite{PhysRevB.1.4555}. The jellium surface under consideration has a value of the Wigner-Seitz radius $r_s = (3/(4\pi n(\boldsymbol{r})))^{1/3}$ equal to $1$. The numerically calculated exact refinement factor is shown together with both the GEs in Fig.~\ref{fig:fig1} (c, d) far inside the surface where $s, q \to 0$ and across the surface region. The results are very similar to Fig.~\ref{fig:fig1} (a, b) for the AG. We see how the KED is accurately described by the AG-GE in Eq.~(\ref{eq:ag_corrections}) in the region of slowly varying electron density, while the ETF GE produces oscillations with a slightly too low amplitude. As we discussed for the AG, also here the ETF GE reproduces the exact behavior better across the surface region, whereas the AG-GE deviates more as we leave the region of slowly varying electron density.

\subsection{The isotropic harmonic oscillator} 

Next we consider a model system that is very different from the AG and jellium surface: the isotropic (radially symmetric) harmonic oscillator (HO) in three dimensions. This model system contains a finite number $N$ electrons. The potential is 
\begin{equation} 
v^{\mathrm{HO}}_{\mathrm{s}}(r) = \frac{1}{2}\omega^2r^2,
\end{equation}
where $\omega$ is the angular frequency and $r$ is the radial distance from the position of equilibrium. We let $\eta = 0,1, \ldots$ be the collective principal quantum number of the three oscillating modes and introduce the curvature parameter $w=\omega/2$. At curvature $\omega$ the $(\eta +1)$\textsuperscript{th} energy level is filled, where
\begin{equation}
\eta = \floor*{\frac{1}{2w} - \frac{3}{2}}.
\end{equation}
Hence, systems with smaller $w$ has a wider potential and contain more electrons. 

\begin{figure*} 
  \begin{tabular}{c @{\hspace{0.05\linewidth}} c}
    \multicolumn{1}{l}{a)} & \multicolumn{1}{l}{b)} \\
    \includegraphics[]{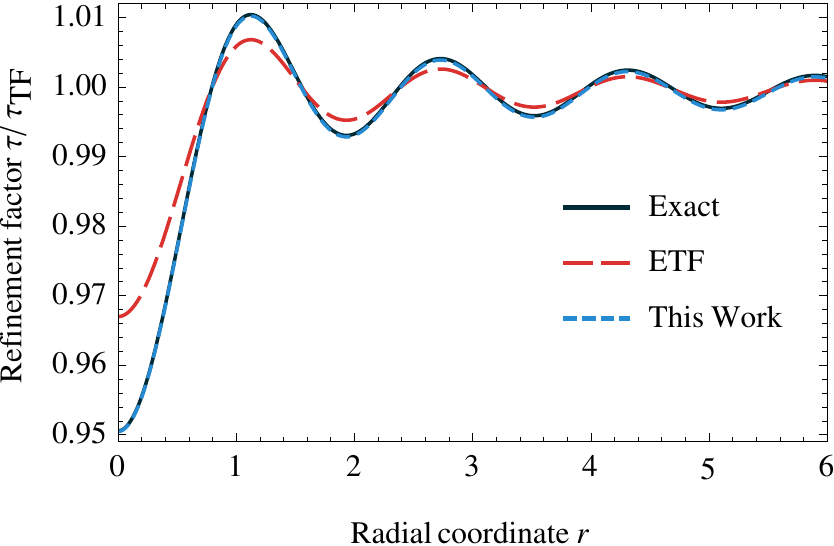} &
    \includegraphics[]{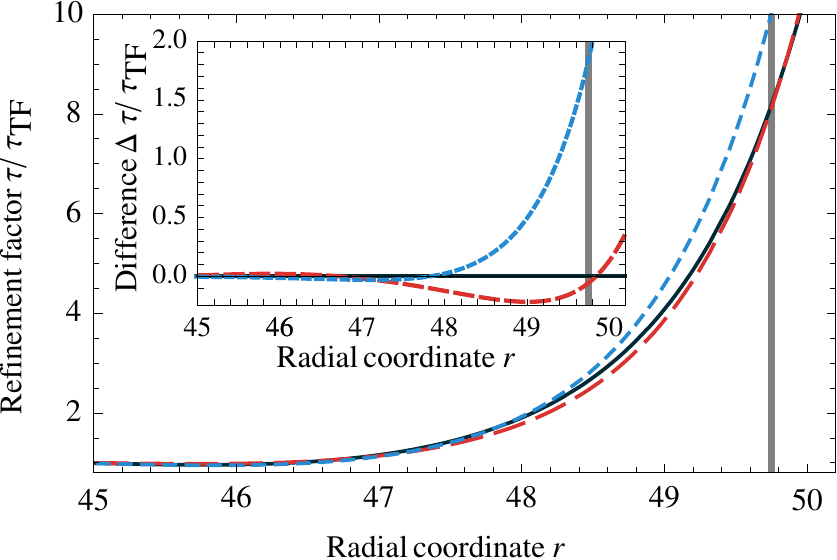} \\
   \multicolumn{1}{l}{c)} & \multicolumn{1}{l}{d)} \\
    \includegraphics[]{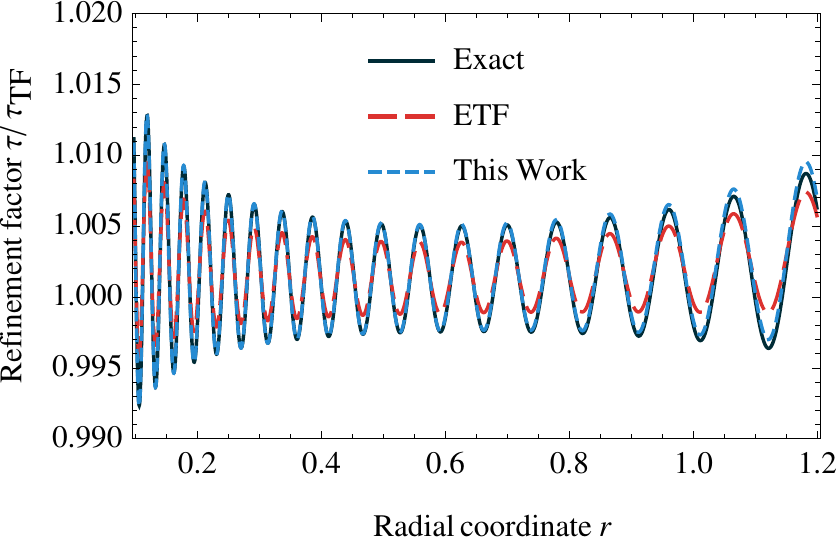} &
    \includegraphics[]{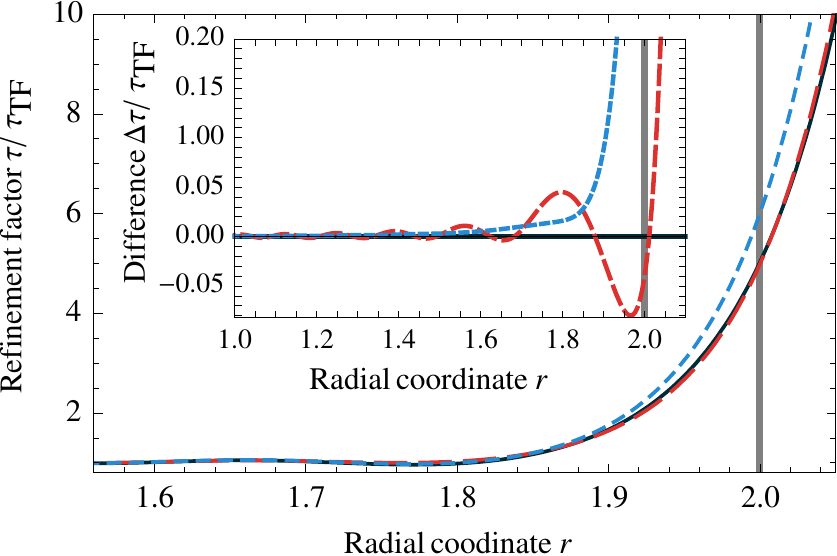} \\
  \end{tabular}
  \caption{The ETF GE and the AG-GE in Eq.~(\ref{eq:ag_corrections}) compared to the exact KED for the isotropic harmonic oscillator (a,b) and the hydrogen-like model system (c,d). Both systems show the same characteristics as was observed in Fig.~\ref{fig:fig1}. The AG-GE describes accurately the oscillations where the density is slowly varying, i.e., $s$ and $q$ are small (shown in panels a and c), whereas ETF fails to do so. However, across the edge (shown in panels b and d), the error in ETF GE is generally smaller than for the AG-GE. 
  } 
  \label{fig:fig2}
\end{figure*}

Figure~\ref{fig:fig2} (a,b) shows a HO system filled up to the $30$\textsuperscript{th} energy level both close to the center (where the density is slowly varying) and across the surface. Despite the fact that the isotropic HO is a closed finite system, the results are surprisingly similar to the open AG and jellium models shown in Fig.~\ref{fig:fig1}. In the regime where the electron density is slowly varying, i.e., where $s$ and $q$ are relatively small, the oscillations reproduced by the ETF are too small in amplitude, but they are well reproduced by the AG-GE in Eq.~(\ref{eq:ag_corrections}). On the other hand, across the surface the ETF follows more closely the exact KED, also for this finite system.


\begin{figure*}
  \begin{tabular}{c @{\hspace{0.05\linewidth}} c}
    \multicolumn{1}{l}{a)} & \multicolumn{1}{l}{b)} \\
    \includegraphics[]{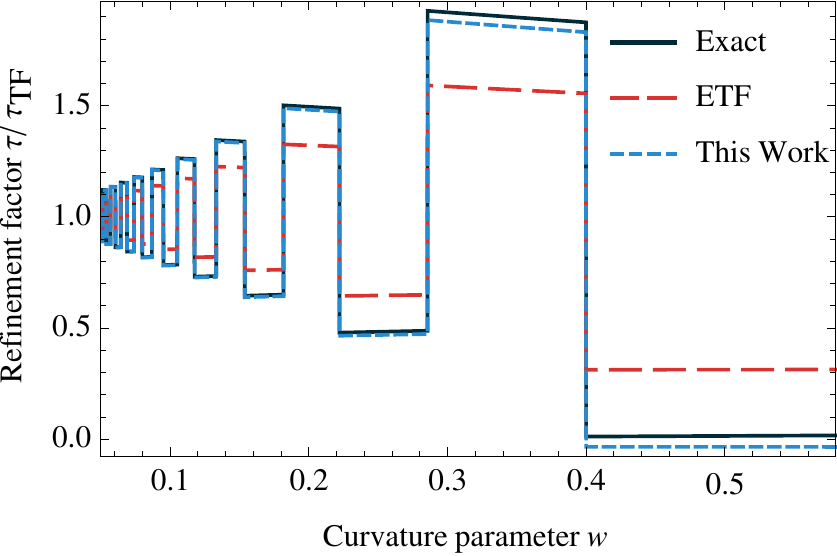} &
    \includegraphics[]{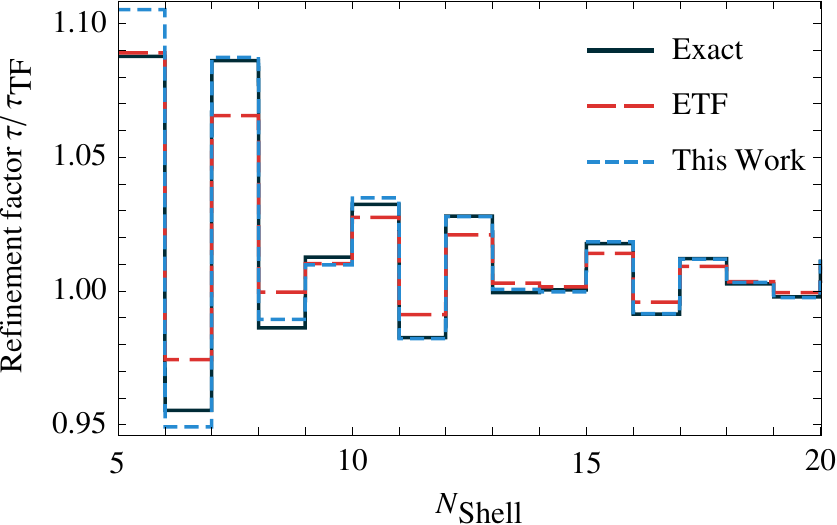} \\
  \end{tabular}
  \caption{Convergence of the ETF GE and the AG-GE in Eq.~(\ref{eq:ag_corrections}) in an arbitrarily chosen single point for the isotropic HO (a) and the hydrogen-like model (b). (a) The isotropic HO at $r_0=0.2$ for the curvature parameter $w \to 0$, i.e., as the system fills up with more particles. Even at moderate values of $w$, the AG-GE reproduces the exact KED with great accuracy, whereas the ETF GE does not. (b) The hydrogen-like model at the arbitrarily chosen point $r_0=0.5$ as the number of filled shells $N_{\mathrm{shell}}$ increases. Similar to (a) the AG-GE accurately reproduces the KED in the limit of large $N_{\mathrm{shell}}$, but the ETF GE does not.} 
  \label{fig:fig4} 
\end{figure*}

The study in Fig.~\ref{fig:fig2} is of a specific highly filled HO system such that the electron density is sufficiently slowly varying. However, to truly realize the \emph{limit} of slowly varying electron density in this system requires taking the curvature parameter $w \to 0$. Hence, in Fig.~\ref{fig:fig4} (a) we have selected one arbitrary spatial point in the system, $r_0 = 0.2$, and show the behavior of both GEs as $w \to 0$. This study confirms the forgoing conclusions: the AG-GE in Eq.~(\ref{eq:ag_corrections}) shows a strongly convergent trend, whereas the ETF appear consistently inaccurate as $w \to 0$. 


As an aside on the topic of the KE in the HO model, we remark that the KE and electron densities of the harmonic oscillator in arbitrary dimensions $d$ has been thoroughly investigated by Brack and van Zyl \cite{brack_simple_2001}. In their paper, they show that the local TF kinetic energy $\tau^{\mathrm{TF}}[n]$ is locally a good approximant to the exact kinetic energy density for any value of $d$, but does not discuss in detail the nature of the gradient corrections to the TF. Moreover, for $d=2$ they verify that $\tau^{\mathrm{TF}}[n]$ gives the exact total energy when integrated. Even though it is anticipated that TF densities are good approximations in the particle number limit $N \rightarrow \infty$, this fact has been proved rigorously for the isotropic harmonic oscillator \cite{brack_harmonically_2003}. 

\subsection{Hydrogen-like model}   

Another radially symmetric finite system of principal interest is particles in a hydrogen-like potential. The potential is
\begin{equation}
v^{\mathrm{HL}}_{\mathrm{s}}(\boldsymbol{r}) = - \frac{Z}{|\boldsymbol{r}|},  
\end{equation}
where $Z$ is the atomic number parameter. For any positive $Z$ this system has infinitely many bound states with eigenvalues
\begin{equation}
\epsilon_{\eta} = -\frac{Z^2}{2}\frac{1}{\eta^2},
\end{equation}
where $\eta = 1,2,\ldots$ is the principal quantum number and there is an $\eta^2$-fold degeneracy in quantum numbers $l$ and $m$. We take the system to be filled with particles up to $\eta = N_{\mathrm{shell}}$, which means that it contains $N = 2\sum_{\eta=1}^{N_{\mathrm{shell}}} \eta^2 = (1/3)N_{\mathrm{shell}}(N_{\mathrm{shell}}+1)(2N_{\mathrm{shell}}+1)$ particles (including the spin degree of freedom). Note that the $Z$ parameter just becomes a scaling factor in all expressions. Hence, the hydrogen-like system is effectively a one-parameter model. There is therefore no need to enforce $Z$ = $N$ (i.e., the choice that would correspond to a neutral atom for interacting particles), and we can instead take $Z=N^2_{\mathrm{shell}}$ to simplify the expressions without loss of generality. This makes the single parameter in this model $N_{\mathrm{shell}}$. Appendix~\ref{hydro-properties} give further details on this model, including the expression for $\tau$. 

Figure~\ref{fig:fig2} (c, d) shows a hydrogen-like model system filled up to the $30$\textsuperscript{th} shell. In Fig.~\ref{fig:fig2} (c) the region where the electron density varies the least is pictured, i.e., where $s$ and $q$ are of smallest magnitude at an intermediate distance from the center. In Fig.~\ref{fig:fig2} (d) a region across the electronic surface of the atom is shown. The hydrogen-like model system shares the same general behavior as the other models considered so far. In the region where the electron density is slowly varying (i.e., where $s$ and $q$ are relatively small), the oscillatory behavior is well described by the AG-GE in Eq.~(\ref{eq:ag_corrections}), whereas the ETF give oscillations with too small amplitude. Across the surface we see how the ETF again follows the exact KED much more closely than the AG-GE. We also provide in Fig.~\ref{fig:fig4} the behavior for the hydrogen-like model of the two GEs for one arbitrarily chosen spatial point in the system, $r_0=0.2$ as $N$ is increased. This test further corroborates the conclusion of the convergence of the AG-GE in the limit of slowly varying density.

\subsection{The Mathieu gas} 

We now turn to a model system that, in contrast to the previous model systems, can model a weakly disturbed uniform electron gas. The model system is the Mathieu Gas (MG), which was investigated in detail in Ref.~\onlinecite{armiento_subsystem_2002}. The potential is taken to be periodic in $z$-dimension
\begin{equation}\label{eq:mg_potential}
v^{\mathrm{MG}}_{\mathrm{s}}(\boldsymbol{r}) = \lambda \boldsymbol{(}1 - \cos(pz)\boldsymbol{)},
\end{equation}
where $\lambda$ is the amplitude and $p$ is the wave vector assigned to the oscillation. We introduce the scaled parameters
\begin{equation}
\bar{\lambda}=\frac{\lambda}{\mu}, \quad \bar{p} = \frac{p}{2k_F}, \quad \text{and} \quad {\bar{z}} = k_F z,
\end{equation}
where $k_F = \sqrt{2\mu} $ is the Fermi wave vector of the uniform electron gas, hence taken in the semi classical limit as independent of position. Depending on the choice of $\bar\lambda$ and $\bar p$, the chemical potential $\mu$ will be above or below the maximums of $v_s$, i.e., the MG can be made to represent either a disturbed uniform electron gas or a system with an infinite number of classically forbidden regions. For further details on the properties relevant to this work see Appendix \ref{mg-properties}. 

\begin{figure}[htb] 
	\centering 
 	\includegraphics[height=0.6\columnwidth, angle=0]{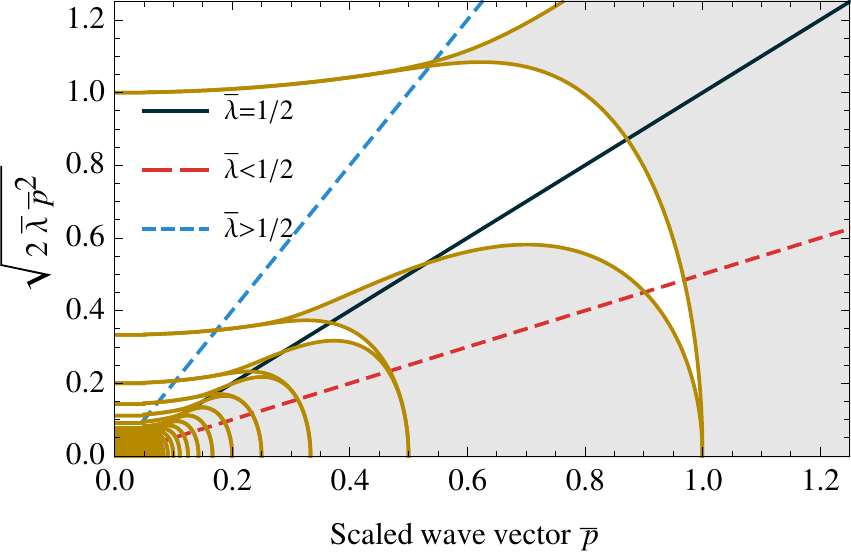}
  	\caption{The parameter space of the Mathieu gas. The shaded area is defined by values of the chemical potential $\mu$ that belong to one of the possible energy bands of the energy spectrum in the $z$-dimension. The light area is defined by values of the chemical potential $\mu$ in the FE continuum in $x$ and $y$ between the bands. Yellow lines correspond to values of the chemical potential situated precisely on the band edges of the energy spectrum in $z$ dimension. The straight lines in the parameter space correspond to different paths along which one can approach the limit of slowly varying electron density in the MG. The black line splits the parameter space into two distinct regions. The blue dashed line indicates a path to the limit of slowly varying density for which $\bar{\lambda}<1/2$, and the red dashed line indicates a path to the limit of slowly varying density for which $\bar{\lambda}>1/2$.}
 	\label{fig:mg-parameter} 
\end{figure}

This model system has a two-dimensional parameter space set by the unitless numbers $\bar{\lambda}$ and $\bar{p}$. When $\bar{\lambda} \rightarrow 0$ we approach the free electron gas and when $\bar{\lambda} \rightarrow \infty$ the occupied energy levels in the $z$-direction reach those of an Hermite gas (HG), see Section \ref{hermite_gas}. The parameter space is shown in Fig.~\ref{fig:mg-parameter}. Any sequence of MG systems that approach the origin of this plot ($\sqrt{2\bar\lambda \bar p^2} \to 0$ and $\bar p \to 0$) represents a possible limit of slowly varying density. There are infinitely many sequences of this kind. A path with $\bar{\lambda}>1/2$ means the chemical potential stays below the maximum of the potential, and the system will have infinitely many classically forbidden regions. A path with $\bar{\lambda}<1/2$ means the chemical potential stays above the maximums of the potential, and the system approach a slowly varying limit along a path that resembles a distrubed uniform electron gas. The path with exactly the border value $\bar{\lambda}=1/2$ represents systems where the chemical potential exactly tangents the maximums of the potential. This path is indicated by a solid line of Fig.~\ref{fig:mg-parameter} and splits the entire parameter space into two distinct regions, which we will refer to as the HG regime (values of $\bar{\lambda}>1/2$) and the free-electron (FE) regime (values of $\bar{\lambda}<1/2$). 


\begin{figure*}
  \begin{tabular}{c @{\hspace{0.05\linewidth}} c} 
    \multicolumn{1}{l}{a)} & \multicolumn{1}{l}{b)} \\
    \includegraphics[]{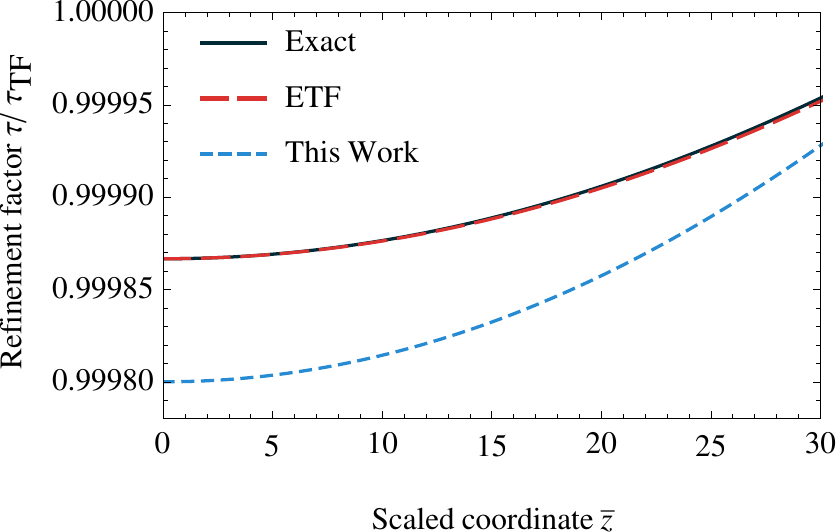} &
    \includegraphics[]{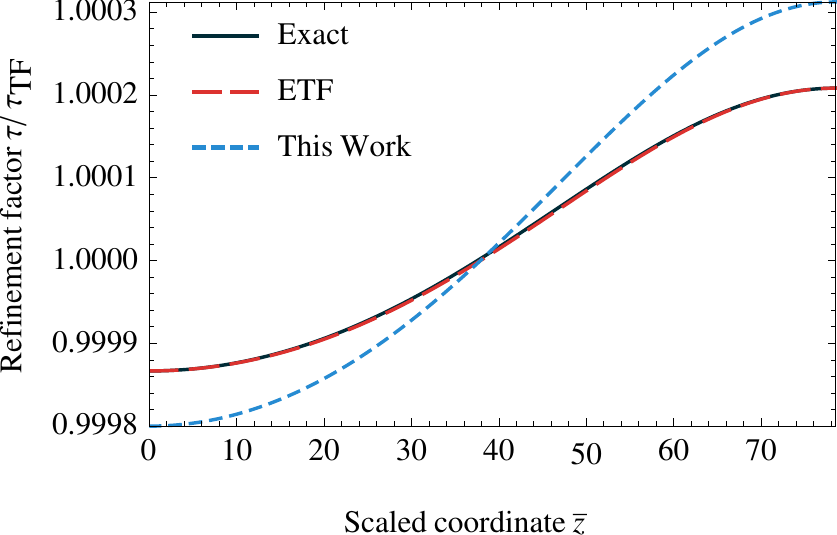} \\
   \multicolumn{1}{l}{c)} & \multicolumn{1}{l}{d)} \\
   \includegraphics[]{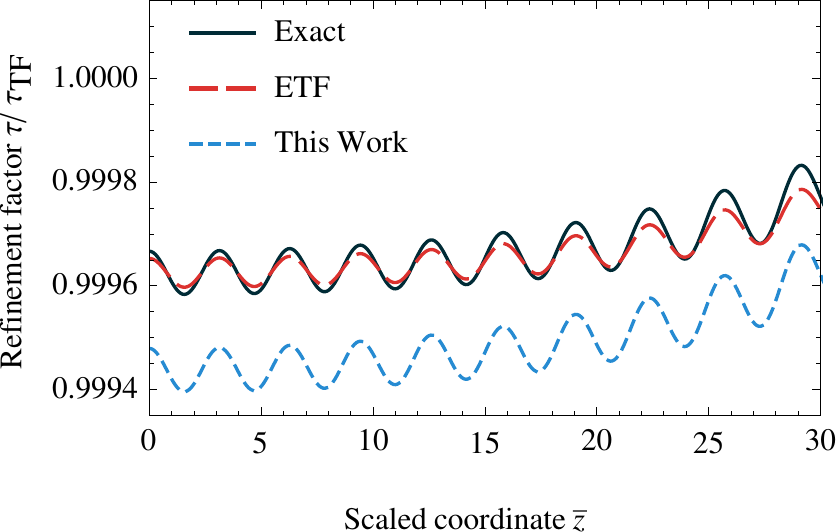} &
    \includegraphics[]{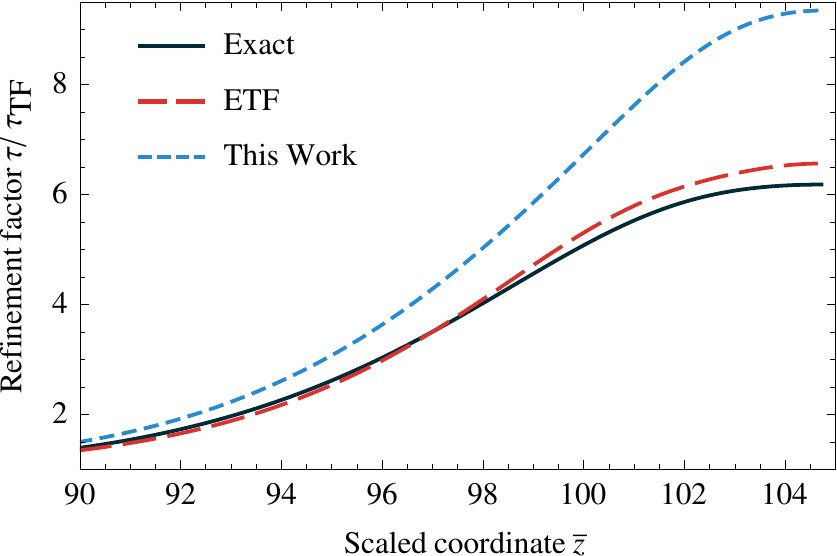} \\
  \multicolumn{1}{l}{e)} & \multicolumn{1}{l}{f)} \\
  \includegraphics[]{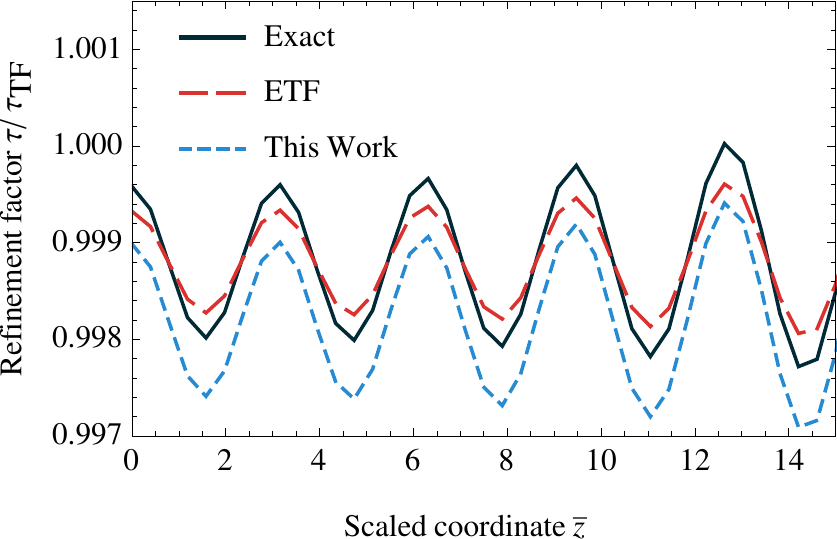} &
   \includegraphics[]{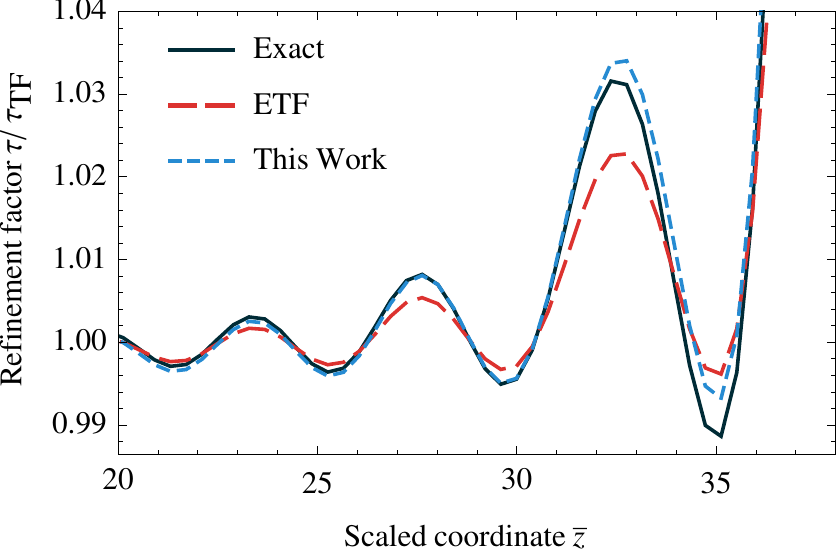} \\  
  \end{tabular}
  \caption{The ETF GE and the AG-GE in Eq.~(\ref{eq:ag_corrections}) compared to the exact KED for three different MG systems as a function of scaled spatial coordinate $\bar{z}$. (a) and (b) The MG with $\bar{\lambda}=0.1,\ \bar p = 0.02$, which has no forbidden regions and thus is FE-like. The ETF GE accurately describes the KED, whereas the AG-GE fails to do so. (c,d) The MG with $\bar{\lambda}=0.5,\ \bar p = 0.015$. In this MG the chemical potential tangents the top of the potential. The ETF GE describes the system fairly well, but there appears to be a minor discrepancy in the amplitude of the oscillations. The AG-GE gives a description that is severely down-shifted as compared to the exact KED. (e,f) The MG with $\bar{\lambda}=0.9,\ \bar p = 0.02$, where the chemical potential is far below the maximum values of the cosine potential, i.e., a system with an infinite number of classically forbidden regions. Both the ETF GE and AG-GE have oscillations that are similar to the exact KED. However, the oscillations in ETF appears to be of too small amplitude, whereas the AG-GE is shifted down relative to the exact KED.\label{fig:fig5}}
\end{figure*}

Our numerical investigation focuses on three different MG systems. The first MG system has $\bar{\lambda}=0.1$ and $\bar p = 0.02$. This MG is in the FE regime of the parameter space, i.e., the system has no classical turning points. In Fig.~\ref{fig:fig5} (a) and (b) we show the KED of this system compared to the ETF GE and the AG-GE as function of the scaled coordinate $\bar{z}$. As should be expected for a system which essentially is a straightforward realization of a weakly disturbed uniform electron gas, the ETF GE describes the KED well. On the other hand, the AG-GE clearly fails to reproduce the KED and appears shifted down even in the region near $\bar z = 0$, i.e., the region near the potential minimum where the values of $s$ and $q$ are the smallest.


Next we consider a MG with $\bar{\lambda}=0.5$ and $\bar p = 0.015$, i.e., $\mu$ is precisely on the intersecting line between the HG and the FE regime. In this system $\mu$ precisely tangents the maximums of the cosine potential. The result is shown in Fig.~\ref{fig:fig5} (c) and (d). In Fig.~\ref{fig:fig5} (c) oscillations have started to form because surface-like behavior starts to manifest even before there are strictly classically forbidden regions-- it is sufficient with regions where the potential maximums are close to the chemical potential. The ETF GE captures these oscillations but there is a visible discrepancy in the amplitude of the oscillations prevalent throughout the system. The AG-GE also displays the oscillations, but with a downwards shift compared to the exact KED. Fig.~\ref{fig:fig5} (d) shows the surface-like region of this MG. By periodicity, these regions repeat throughout the entire system. Similar to the other model systems studied above, the ETF appear to better reproduce the behavior in this region.

We finally consider the MG with an amplitude $\bar{\lambda}=0.9$ and $\bar p = 0.02$. Here, $\mu$ is far below the the maximums of the cosine potential. The exact KED is compared to the ETF GE and the AG-GE in Figs.~\ref{fig:fig5} (e) and (f). The general features are similar to the $\bar{\lambda}=0.5$ case but more clear here. Both the GEs display oscillations similar to those in the exact KED. The ETF GE appears to underestimate the amplitude, whereas the AG-GE is shifted down. In absolute numbers, the incorrect offset of the AG-GE increases along the sequence of systems with increasing amplitude $\bar \lambda$, shown in Fig.~\ref{fig:fig5} subfigures (a)$\to$(c)$\to$(e). However, relative to the size of the oscillations that appear in the systems, the offset becomes smaller.

\begin{figure*}
  \begin{tabular}{c @{\hspace{0.05\linewidth}} c}
    \multicolumn{1}{l}{a)} & \multicolumn{1}{l}{b)} \\
    \includegraphics[]{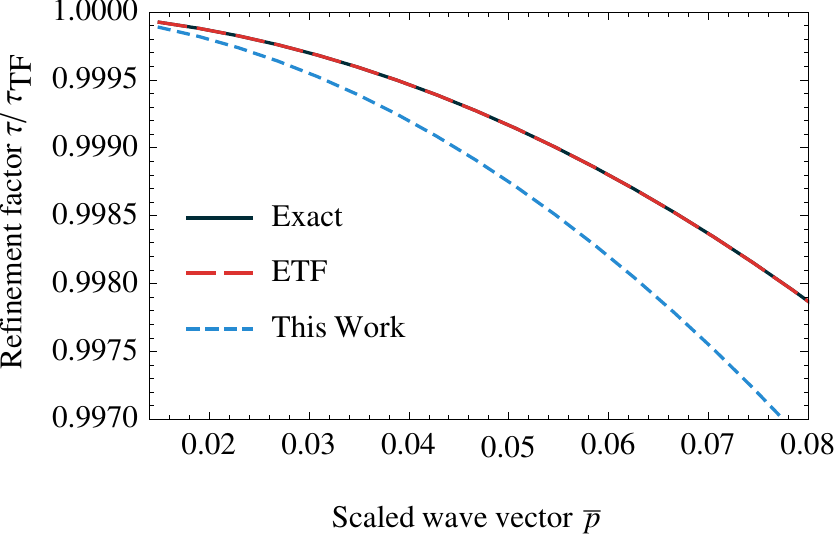} &
    \includegraphics[]{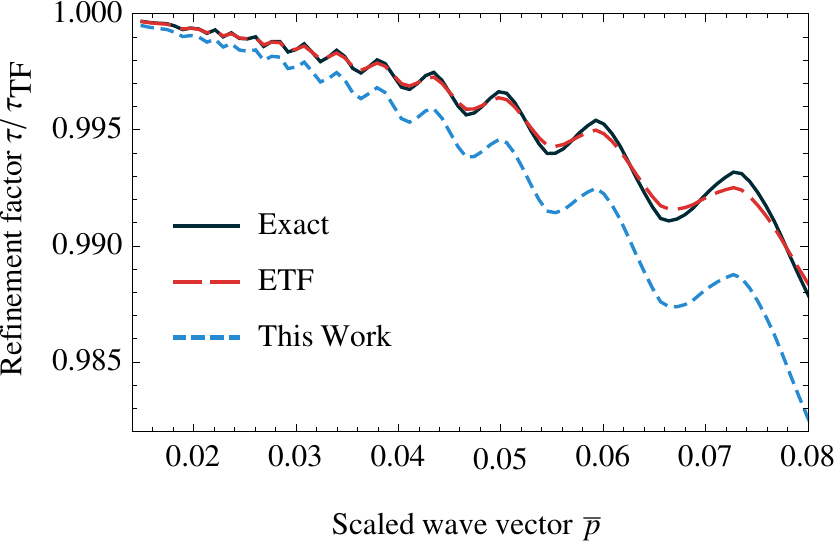} \\
    \multicolumn{1}{l}{c)} & \multicolumn{1}{l}{d)} \\
    \includegraphics[]{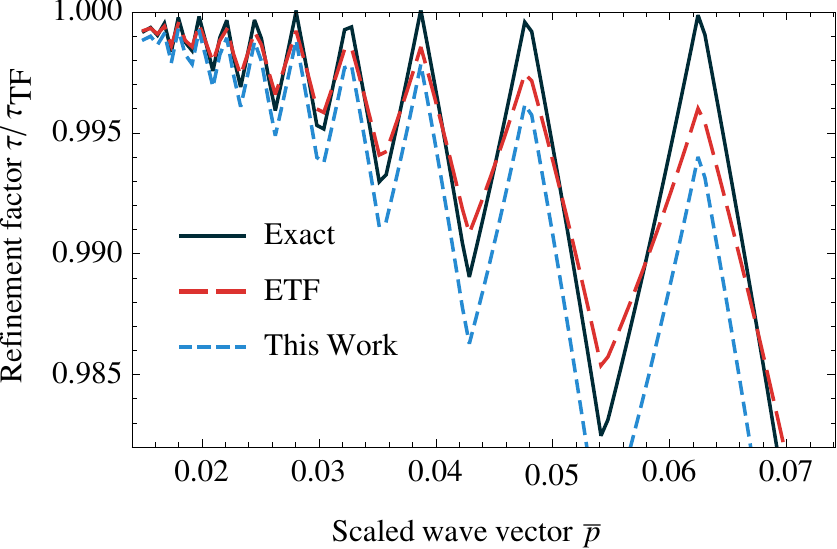} &
    \includegraphics[]{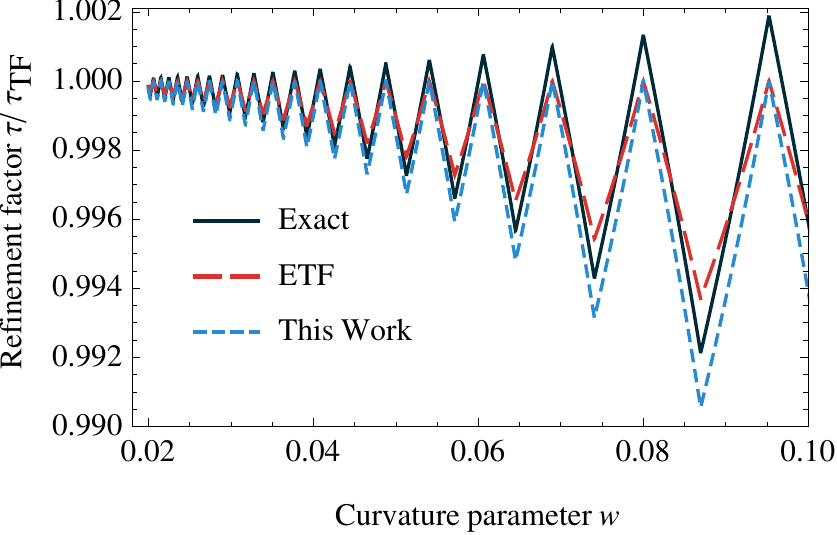} \\
  \end{tabular}
  \caption{Convergence of the ETF GE and the AG-GE in Eq.~(\ref{eq:ag_corrections}) in one single arbitrarily chosen spatial point for MG systems ($\bar z = 0.01$) with different parameters (a-c) and in the HG for $\bar z = 0$ (d). (a) The convergence as the scaled wave vector $\bar{p}$ approaches zero for a MG with $\bar{\lambda}=0.1$. In this MG there are no classical turning points and the ETF GE appears to reproduce the exact KED well, whereas the AG-GE generally fails. (b) A MG with $\bar{\lambda}=0.5$ for which the chemical potential tangents the top of the potential. (c) A MG with $\bar{\lambda}=0.9$ where the chemical potential is far below the maximum values of the cosine potential, i.e., a system with an infinite number of classical turning points. Both the ETF GE and the AG-GE have oscillations that are similar to the exact KED in (b) and (c). However, the oscillations in ETF are of too small amplitude, whereas the AG-GE appears to be shifted down. (d) The HG in the limit of a curvature parameter $w \to 0$. The general behavior in the MG with large $\bar\lambda$ in panel (c) and the HG in panel (d) are very similar.} 
 \label{fig:fig3}
\end{figure*}

We now move on to compare the convergence between the ETF GE and the AG-GE for one single arbitrary spatial point of the MG. For the three different MG systems with the rescaled amplitudes $\bar{\lambda}=0.1$, $\bar{\lambda}=0.5$ and $\bar{\lambda}=0.9$ we will study the behavior in the arbitrarily chosen spatial point at scaled coordinate $\bar{z}=0.01$. We let the scaled wave vector $\bar{p}$ approach zero which takes us in a limit of a slowly varying density, $s, q \to 0$. The result is shown in Fig.~\ref{fig:fig3}. In Fig.~\ref{fig:fig3} (a) the MG with $\bar{\lambda}=0.1$ is shown. As $\bar{p} \rightarrow 0$, the ETF GE moves closer to the exact KED, whereas the AG-GE consistently is too low. Fig.~\ref{fig:fig3} (b) shows the convergence when $\bar{\lambda}=0.5$. Now, the KED curve has visible oscillations. Then, as we approach the high amplitude limit with $\bar{\lambda}=0.9$ we see in Fig.~\ref{fig:fig3} (c) how the oscillations become sharper. The Figs.~\ref{fig:fig3} (b) and (c) confirm the same conclusions as was found in Fig.~\ref{fig:fig5}: neither the AG-GE, nor ETF, appears to describe the limit of slowly varying density in MG model systems with large values of $\bar{\lambda}$ well. The ETF reproduces the oscillations with too small amplitude, and Eq.~(\ref{eq:ag_corrections}) is generally shifted down with respect to the exact KED.


\subsection{The Hermite Gas}\label{hermite_gas}  

Since we identified in the previous subsection an issue with describing the KED of the MG in the large amplitude regime i.e., $\bar{\lambda}>1/2$, we will now study a model system that represents the extreme case of $\bar{\lambda} \rightarrow \infty$. This limit is an EG with an harmonic oscillator potential, the Hermite Gas (HG) discussed in Refs.~\onlinecite{armiento_subsystem_2002, PhysRevB.82.115103}. The potential is
\begin{equation}
v^{\mathrm{HG}}_{\mathrm{s}}(z) = \frac{\omega^2}{2} z^2
\end{equation}
where $\omega$ is the angular frequency of the oscillation and $z$ is the distance from the equilibrium. The normalized eigenfunctions in the $z$-direction are proportional to Hermite polynomials. The scaled parameters in this case are 
\begin{equation}
w= \frac{\omega}{\mu},\ N(\mu) = \floor*{ \frac{1}{w} - \frac{1}{2}},\ \bar{z}=k_F z,
\end{equation}
where $k_F = \sqrt{2 \mu}$, $\mu$ is the chemical potential and $N(\mu)$ is the number of occupied orbitals in the $z$-direction which is directly linked to the \emph{curvature parameter} $w$ which sets the curvature, and thus the width, of the potential parabola. Hence, the HG is effectively a one-parameter model. See Ref.~\onlinecite{PhysRevB.82.115103} for additional properties of the HG. Further details regarding the HG relevant to this work can be found in Appendix \ref{hg-properties}. The limit of slowly varying density is reached for $w \rightarrow 0$, which opens the harmonic potential to infinite width, while keeping the chemical potential constant. 

We note briefly that the HG system expressions can be rescaled to give a different, but equivalent, view. The curvature $w$ can be taken as fixed, and the single parameter taken to be $\mu$. The limit of slowly varying density is then realized in a fixed energy spectra of the HG, with the levels filling up as $\mu \to \infty$. 

\begin{figure*}
  \begin{tabular}{c @{\hspace{0.05\linewidth}} c}
    \multicolumn{1}{l}{a)} & \multicolumn{1}{l}{b)} \\
    \includegraphics[]{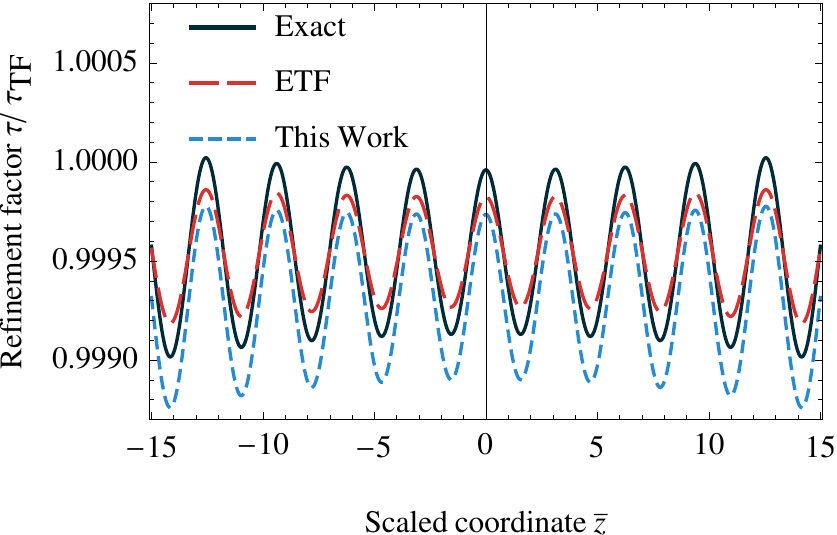} &
    \includegraphics[]{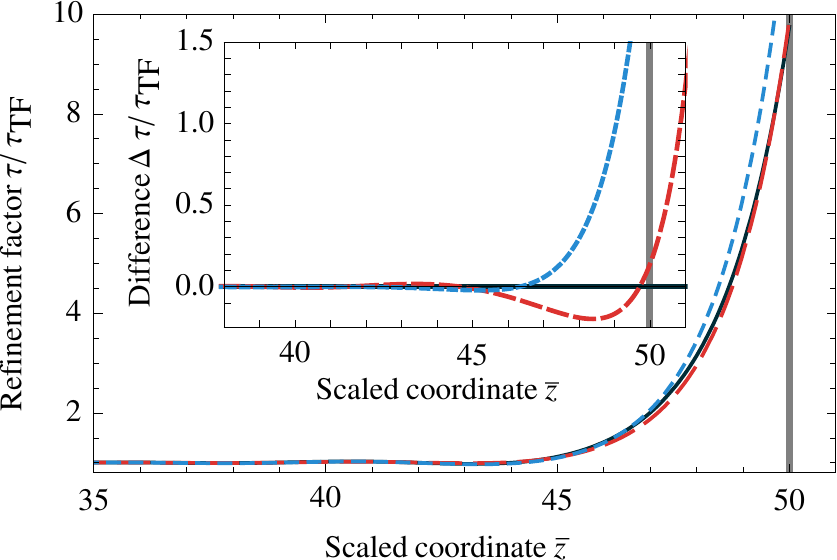} \\
  \end{tabular}
  \caption{The HG filled up to the $30$\textsuperscript{th} energy level in the $z$-dimension energy spectra. (a) The exact KED for this system compared to the ETF GE and the AG-GE in Eq.~(\ref{eq:ag_corrections}) as function of the scaled coordinate $\bar z$. The ETF GE underestimates the amplitude of the exact oscillation, whereas the AG-GE has a relative offset compared to the amplitude. (b) The surface region of the same system, where the vertical line is the classical turning point. The AG-GE has a larger errors than the ETF GE in this surface region.}  
  \label{fig:fig7}
\end{figure*}

We now study a HG filled with with electrons up to the $30$\textsuperscript{th} energy level in the $z$-direction. We are interested first in the classically allowed region. Figure~\ref{fig:fig7} (a) shows the exact KED for this system far away from the classical turning points compared to the ETF GE and the AG-GE in Eq.~(\ref{eq:ag_corrections}) as functions of the scaled coordinate $\bar z$. Similar to our observations for the large $\bar \lambda$ region of the MG model, we find that the ETF GE consistently underestimates the amplitude of the oscillations of the KED. On the other hand, Eq.~(\ref{eq:ag_corrections}) reproduces the amplitude well, but is shifted down compared to the exact KED. The surface region of this system is shown in Fig.~\ref{fig:fig7} (b). The ETF GE and the AG-GE both follow closely the exact KED in this region, but Eq.~(\ref{eq:ag_corrections}) deviates more than ETF near the classical turning point. In Fig.~\ref{fig:fig3} (d) convergence at the chosen point $\bar z = 0$ is shown. The tendency of the ETF GE of underestimating the amplitude of the oscillation and the relative offset of the AG-GE is seen here as well. 


\section{Discussion} 

In this work we have considered the influence of quantum oscillations due to electronic edges in the KED in limit of slowly varying electron density. This has been achieved using the AG model system which is a useful approximation of an inhomogeneous system of many electrons with a well defined surface region\cite{kohn_edge_1998}. Quantum oscillations in the KE and electron densities have previously been discussed \cite{kohn_quantum_1965, thorpe_thouless} and they are known to not be captured by the ETF GE. On the other hand, the AG-GE in Eq.~(\ref{eq:ag_corrections}) is a gradient expansion that includes quantum gradient corrections from a linear surface. It is shown to describe the oscillations in the regime of slowly varying density in the jellium surface model and also in regions of slowly varying density in finite systems, i.e., the isotropic HO and the hydrogen-like model. However, the AG-GE does not appear to represent the KED well in the limit of slowly varying density of the MG and HG model systems. In the low amplitude regime (i.e., $\bar{\lambda}<1/2$) of the MG, this is perfectly expected, since there is no classically forbidden region in the system, i.e., the system has no surface. For this case the ETF GE is within its domain of validity throughout the system without any quantum corrections. One may at a first think that the KED of the \emph{high amplitude} MG, and indeed the HG, should be well represented by the AG-GE, since these systems have surface regions. What is seen however, is that none of the expansions reproduce well the KED in these systems. 

The failure of both the GEs for the MG and HG systems may be the result of system dependence of the quantum corrections in spite of the apparent generality of the AG-GE for other systems with surfaces. However, our results lead us to speculate that the behavior may rather be a consequence of the very strong anisotropy between the dimensions in the MG and HG systems. The energy levels in the AG are infinitely dense in all three dimensions. The two finite systems considered (the HO and the hydrogen-like potential) are spherically symmetric and thus give a finite number of energy levels that are equally dense in all dimensions. However, the MG with large $\bar \lambda$ and the HG both have infinitely dense levels only in the $x$- and $y$-dimensions, whereas for the $z$-dimension there is a discrete spectrum (HO) or a band structure (MG) where the bands get thinner as the potential amplitude $\bar \lambda$ increases. Our hypothesis thus means that the MG and HG systems may display a mix of both the GEs. The ETF describes the changes in the KED as the chemical potential moves \emph{between} two discrete energy levels in the $z$-dimension, since such a change only adds plane-wave states in the $x$- and $y$-dimensions. On the other hand, as the chemical potential moves past the eigenvalues of states in the $z$-dimension, the change in the KED is supposedly described well by the AG-GE. This would explain why the true behavior of the KED in these systems appear to share features of both GEs. However, if this interpretation is true, the AG-GE may be of very broad general validity in describing quantum oscillations in systems with surface regions, as long as the systems do not have unreasonably anisotropic dimensions. 

\subsection{Development of kinetic energy functional $T_s[n]$}  

It is a long term goal of functional development to construct viable approximations of the exchange-correlation and kinetic energy in DFT. Prior work on quantum corrections, as well as the present one, suggest that the electron density and KED in a classically allowed region where the electron density is slowly varying depend on the influence of non-local information, in the sense that the quantum oscillations are determined by the behavior of $v_{\mathrm{s}}({\boldsymbol{r}})$ outside the immediate neighborhood of the point $\boldsymbol{r}$. Surfaces and/or classical turning points are examples of such signatures in the topological landscape of the single particle potential that influences the classically allowed region. The local energetics of the system in this region thus appears to differ between dissimilar systems and cannot uniquely be described in terms of a straightforward extension of TF-theory, solely by adding more terms in the local expansion (as is sometimes suggested, see, e.g., Ref.~\onlinecite{pearson_symptotoc}).  

One may at this point ask if a successful general semi-local approximation to $\tau$ valid for all limits of slowly varying electron density is even possible. Is it possible for such a semi-local approximation to differentiate between a situation where the ETF GE applies vs. when the AG-GE applies, based only on the semi-local information available in $n(\boldsymbol{r})$, $s$ and $q$? To investigate this question we look at the path in countor plot of $s^2$ and $q$ as we take the limit of slowly varying density in a few different model systems. 

\begin{figure}[htb] 
	\centering
 	\includegraphics[height=0.6\columnwidth, angle=0]{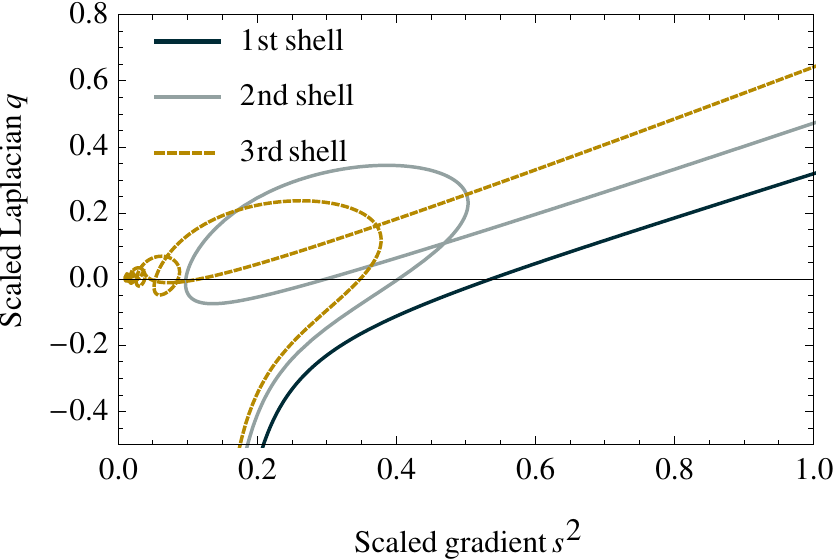} 
  	\caption{Coutour plot that shows $s^2$ vs. $q$ over all values of $r$ for a few hydrogen-like model systems filled up to and including the 3\textsuperscript{rd} shell. As the shell variable $N_{\mathrm{shell}}$ is increased, the number of loops increases and the curve approaches the origin, which demonstrates how the limit of slowly varying density is reached in this system within an intermediate region of $r$ values.}
 	\label{fig:svsq-hydrogen}
\end{figure}

\begin{figure}[htb] 
	\centering
 	\includegraphics[height=0.6\columnwidth, angle=0]{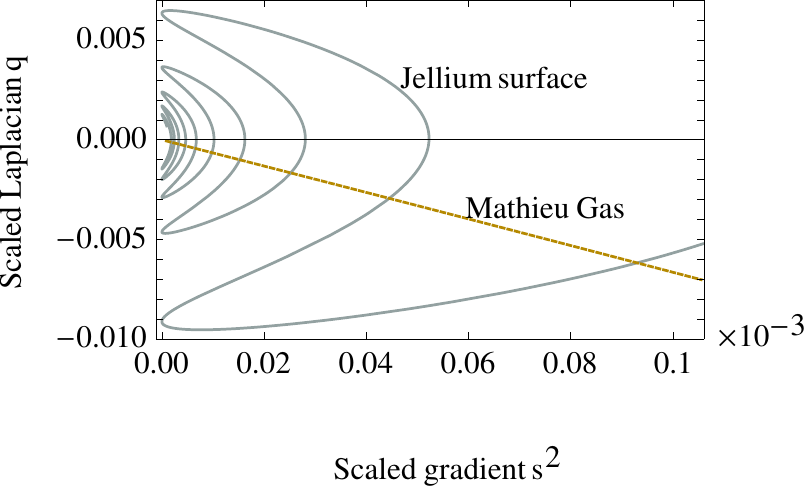} 
  	\caption{Countour plot of the scaled gradient $s^2$ versus the scaled Laplacian $q$ in the limit of slowly varying density for the case of a jellium surface ($r_s=1$) as $z \to -\infty$ and a Mathieu gas ($\bar{\lambda}=0.1$) as $\bar p \to 0$.}
 	\label{fig:jellium-mg} 
\end{figure}

Two such countor plots are shown in Figs.~\ref{fig:svsq-hydrogen} and \ref{fig:jellium-mg}. In Fig.~\ref{fig:svsq-hydrogen} the behavior of the hydrogen-like model is shown to illustrate the complexity of how the limit of slowly varying density is achieved in these model systems. However, to address the question of the information available from $s$ and $q$ alone, we show in Fig.~\ref{fig:jellium-mg} both the jellium surface model and a low amplitude MG overlaid into the same figure. We know that the limit of slowly varying density of the jellium surface model is accurately described by the AG-GE, whereas the ETF GE describes this particular low amplitude MG very well. As is seen in the figure, the two systems approach the limit of slowly varying density very differently. However, there are points where the curves intersect and the local value of the two expansions are not the same in these points. This suggests that there cannot exist a simple semi-local approximation of $F(s, q)$ that gets both types of limits right. This appear to be a problematic conclusion in the development of approximative expressions. Nevertheless, it is possible that the precise difference between the AG-GE and the ETF GE this far into the limit of slowly varying electron density turns out to be energetically less relevant than other features of the KED.

\section{Summary and Conclusions}

We have studied the AG in the limit far inside the surface where the electron density is slowly varying. The presence of a surface region in the system (i.e., where the density decays to zero) requires quantum corrections compared to the usual ETF GE. We have derived an expression for a GE incorporating such quantum corrections from the AG, the AG-GE, and find it to describe systems of both finite and infinite size with surface regions well. However, neither ETF nor the AG-GE appear to apply directly to the two model systems considered where the energy level spacing is very anisotropic between different dimensions. 

Furthermore, while the present work has exclusively discussed the kinetic energy $\tau$, we note that the gradient coefficient in the GE for the exchange energy has been subject of some debate, with Kleinman and Lee \cite{PhysRevB.37.4634} arriving at the presently accepted value of $10/81$ for a partially integrated GE that avoids a Laplacian term. The present work highlights the question if an alternative local exchange energy density GE may exist that takes into account the quantum corrections due to a surface region. However, the situation for a local GE of the exchange energy density is much less clear than for the KED. Two of us have, in previous works, found that even for the MG in the limit of a weakly perturbed uniform electron gas, no such local GE appear to exist \cite{armiento_subsystem_2002, PhysRevB.68.245120}. 

\begin{acknowledgments} 

We acknowledge support from the Swedish Research Council (VR), Grant No. 621-2011-4249 as well as the Linnaeus Environment at Link\"oping on Nanoscale Functional Materials (LiLi-NFM) funded by VR. A.L. thanks Dr. Olle Hellman for valuable input on the presentation.
\end{acknowledgments} 

\appendix \label{radial}

\section{Systems with radial symmetry}\label{radial} 

Consider $N$ non-interacting fermions in the spherically symmetric potential $v_{\mathrm{s}}(r)$. The separable solutions to $\hat{H}_{\mathrm{s}} \phi_{\nu} = \epsilon_{\nu} \phi_{\nu}$ are
\begin{equation}
\phi_{\nu} = \phi_{nlm}(r,\theta,\phi)=R_{nl}(r)Y_{lm}(\theta,\phi),
\end{equation}
where each $R_{nl}$ is a radial distribution function and $Y_{lm}$ are spherical harmonics. We wish to calculate the positive kinetic energy density
\begin{equation}
\tau(r,\theta,\phi) = \sum_{nlm}|\nabla \phi_{nlm}|^2.
\end{equation}
To this end, we calculate the gradient $\nabla \phi_{nlm}$ using the product rule
\begin{equation} 
\nabla \phi_{nlm} = \nabla R_{nl} Y_{lm} + R_{nl} r \nabla Y_{lm}  
\end{equation}
where $\boldsymbol{\Psi}_{lm}=r \nabla Y_{lm}$ is the vector spherical harmonics along the $\phi$-direction. The $\boldsymbol{\Psi}_{lm}$ obey Uns\"{o}ld's theorem\cite{unsold_theorem} i.e.,
\begin{equation}
\sum_{m=-l}^{l}| \boldsymbol{\Psi}_{lm} | ^2 = \frac{1}{4 \pi}(2l+1)(l+1)l,
\end{equation}
which mirrors the fact that spatial densities must be radially symmetric. In the same way, the $Y_{lm}$'s obey the well known addition theorem i.e.,
\begin{equation}
\sum_{m=-l}^{l}| Y_{lm} | ^2 = \frac{1}{4 \pi}(2l+1)
\end{equation}
Hence, we are left with the expression for the radial KED:
\begin{eqnarray}\label{eq:tauspherical}
\tau^{\mathrm{Radial}}(r) &=& \frac{1}{4\pi} \sum_{n=1}^{N} \sum_{l=0}^{n-1}\big(2l+1\big) \nonumber \\ 
&\times& \left(\left[\frac{\partial}{\partial r} R_{nl}(r)\right]^2 + l(l+1)\left[\frac{R_{nl}(r)}{r}\right]^2\right).
\end{eqnarray}

\subsection{KED of the hydrogen-like atom}\label{hydro-properties} 

Consider non-interacting electrons that are bound by the hydrogen-like potential 
\begin{equation} 
v^{\mathrm{HL}}_{\mathrm{s}}(\boldsymbol{r}) = - \frac{Z}{|\boldsymbol{r}|}, 
\end{equation}
where $Z$ is the atomic number. The system consists of a finite number of $N$ electrons and the KS orbitals are the familiar functions 
\begin{equation}
\phi_{\eta lm}(r,\theta,\phi) = R_{\eta l}(r)Y_{lm}(\theta,\phi)
\end{equation}  
where the $R_{\eta l}$'s are proportional to Laguerre polynomials and the $Y_{lm}$'s are the normalized spherical harmonics. The eigenvalues are
\begin{equation}
\epsilon_{\eta} = -\frac{Z^2}{2}\frac{1}{\eta^2} .
\end{equation}
We take the system to be filled with particles up to principal quantum number $\eta = N_{\mathrm{shell}}$, which means that it contains $N = 2\sum_{\eta=1}^{N_{\mathrm{shell}}} \eta^2 = (1/3)N_{\mathrm{shell}}(N_{\mathrm{shell}}+1)(2N_{\mathrm{shell}}+1)$ particles (including the spin degree of freedom). Furthermore, as explained in the main text, the atomic number $Z$ scales the system, so we set $Z=N_{\mathrm{shell}}^2$ to simplify the problem. Using Eq.~(\ref{eq:tauspherical}) the positive KED $\tau$ becomes
\begin{eqnarray}
\tau^{\mathrm{HL}}(r) &=& \frac{1}{4\pi} \sum_{\eta=1}^{N_{\mathrm{shell}}} \sum_{l=0}^{\eta-1}(2l+1) \left( \frac{2Z}{\eta} \right)^3 \frac{(\eta-l-1)!}{2\eta[(\eta+l)!]} \nonumber \\
&\times& \bigg( \left[\frac{\partial}{\partial r} \left(e^{-\frac{Zr}{\eta}}\left( \frac{2Zr}{\eta}\right)^lL_{\eta-l-1}^{2l+1}\left( \frac{2Zr}{\eta}\right) \right) \right]^2 \nonumber \\
&+& l(l+1) \left[\frac{1}{r}e^{-\frac{Zr}{\eta}}\left( \frac{2Zr}{\eta}\right)^lL_{\eta-l-1}^{2l+1}\left( \frac{2Zr}{\eta}\right) \right]^2 \bigg). \nonumber \\
\end{eqnarray}

\section{Properties of the Mathieu gas}\label{mg-properties} 

Given a potential which is constant in two spatial directions and varies along the third, $z$ say, according to
\begin{equation}
v^{\mathrm{MG}}_{\mathrm{s}}(z) = \lambda \boldsymbol{(}1 - \cos(pz)\boldsymbol{)},
\end{equation}    
where $\lambda$ is the amplitude and $p$ is the wave vector of the oscillation, the solutions to the corresponding eigenvalue problem are of the form
\begin{eqnarray}
\varphi^{\mathrm{MG}}_{\eta}(z) &=& \frac{1}{L_3} \left[ ce_{\eta} \left(z, -\frac{1}{2} \frac{\bar{\lambda}}{\bar{p}^2} \right) + ise_{\eta} \left(z, -\frac{1}{2}\frac{\bar{\lambda}}{\bar{p}^2} \right)  \right] \nonumber \\
&=& \frac{1}{L_3}e^{i \eta \bar{p}\bar{z} } \sum_{k \in \mathbb{Z}}c_{2k}^{\eta}e^{i2k \bar{p}\bar{z}},
\end{eqnarray} 
where $\eta \bar{p} k_F L_3=2\pi n_3$ ($n_3 \in \mathbb{Z}$), $L_3$ is the size of the system, measured in units of $z$. Here, the functions $ce_{\eta}$ and $se_{\eta}$ are the real even and odd Mathieu functions respectively. We have introduced the scaled parameters
\begin{equation}
\bar{\lambda}=\frac{\lambda}{\mu}, \quad \bar{p} = \frac{p}{2k_F}, \quad \bar{z} = k_F z,
\end{equation}
where $k_F = \sqrt{2\mu}$ is the Fermi wave vector of the uniform electron gas. 
The $c_{2k}^{\eta}$ are found from the relation
\begin{equation}
(2k + \eta)^2 c_{2k}^{\eta} - \frac{\bar{\lambda}}{2\bar{p}}(c_{2k-2}^{\eta} + c_{2k+2}^{\eta}) = a\left(\eta, \frac{\bar{\lambda}}{2\bar{p}^2} \right)c_{2k}^{\eta},
\end{equation}
and they are normalized according to 
\begin{equation} 
\sum_{k \in \mathbb{Z}}|c_{2k}^{\eta}|^2 = 1.
\end{equation}
The eigenvalue associated with the $\varphi_{\eta}$ are
\begin{equation}\label{eq:eig} 
\frac{\epsilon_{\eta}}{\mu} =\bar{\lambda} + \bar{p}^2a\left( \eta, \frac{\bar{\lambda}}{2\bar{p}^2} \right).
\end{equation} 
Using relation Eq.~(\ref{eq:ked_eg}), the positive kinetic energy density becomes
\begin{eqnarray}
\frac{\tau^{\mathrm{MG}}(z)}{\tau_u} &=& \frac{5}{2}\bar{p}\int_{0}^{\eta_m} d\eta \, \bigg[ \frac{1}{2}\left(1 - \bar{p}a_{\eta}-\bar{\lambda}\right)^2 \nonumber \\
&\times& \left( ce_{\eta}^2(\bar{p}\bar{z}, \bar{q}) + se_{\eta}^2(\bar{p}\bar{z}, \bar{q}) \right) + \bar{p}^2(1 - \bar{p}^2a_{\eta} - \bar{\lambda}) \nonumber \\
&\times& (ce_{\eta}^{\prime 2}(\bar{p} \bar{z}, \bar{q}) + se_{\eta}^{\prime 2}(\bar{p} \bar{z}, \bar{q})) \bigg], 
\end{eqnarray}
where $\tau_u = k_F^5/(10\pi^2)$ and $a_{\eta}$ are the eigenvalues of Eq.~(\ref{eq:eig}), $\eta_m$ is the energy of the highest occupied state, and $\bar{q} = -(1/2) {\bar{\lambda}}/{\bar{p}^2}$. 

\section{KED of the Hermite gas}\label{hg-properties} 

For the HG, the potential varies along the $z$-axis as an HO, i.e., 
\begin{equation}
v^{\mathrm{HG}}_{\mathrm{s}}(z) = \frac{\omega^2}{2} z^2.
\end{equation}
The eigenfunctions are the familiar 
\begin{equation}
\varphi^{\mathrm{HG}}_{\eta}(z) = \left( \sqrt{\frac{\omega}{\pi}} \frac{1}{2^{\eta}\eta!} \right)^{1/2} H_{\eta}\left(\sqrt{\omega}z\right)e^{-\frac{\omega z^2}{2}}, 
\end{equation}
with corresponding eigenvalues
\begin{equation}
\epsilon_{\eta} = \omega \bigg( \eta + \frac{1}{2} \bigg),
\end{equation}
for $\eta = 0,1,\ldots$. Introducing the scaled parameters 
\begin{equation}
w= \frac{\omega}{\mu}, \quad N(\mu) = \floor*{ \frac{1}{w} - \frac{1}{2}}, \quad \bar{z}=k_Fz,
\end{equation}
where $k_F = \sqrt{2 \mu} $, and $N(\mu)$ is the number of occupied $z$-orbitals respectively. Division with the KED of the free electron gas yields the dimensionless quantity
\begin{eqnarray} 
\frac{\tau^{\mathrm{HG}}(\bar{z})}{\tau_u} &=& \frac{1}{4\pi} \sqrt{ \frac{w}{2\pi} } \sum_{\eta = 0}^{N(\mu)} \frac{1}{2^{\eta} \eta!}\left[1 - w\left(\eta + \frac{1}{2}\right) \right] \nonumber \\
&\times& \bigg[ \frac{1}{2}H_{\eta}^2\left(\sqrt{\frac{w}{2}} \bar{z} \right) e^{-\frac{1}{2}w \bar{z}^2} \left(1 - w\left(\eta + \frac{1}{2}\right) \right) \nonumber \\ 
&+& \left[ \frac{d}{d\bar{z}}\left( H_{\eta}\left( \sqrt{\frac{w}{2}} \bar{z} \right)e^{-\frac{\omega \bar{z}^2 }{4}} \right) \right]^2 \bigg].
\end{eqnarray} 
Note that the \emph{curvature parameter} $w$ describing the wideness of the potential parabola now directly determines the number of occupied orbitals in the $z$-direction.  


\begin{thebibliography}{34}%
\makeatletter
\providecommand \@ifxundefined [1]{%
 \@ifx{#1\undefined}
}%
\providecommand \@ifnum [1]{%
 \ifnum #1\expandafter \@firstoftwo
 \else \expandafter \@secondoftwo
 \fi
}%
\providecommand \@ifx [1]{%
 \ifx #1\expandafter \@firstoftwo
 \else \expandafter \@secondoftwo
 \fi
}%
\providecommand \natexlab [1]{#1}%
\providecommand \enquote  [1]{``#1''}%
\providecommand \bibnamefont  [1]{#1}%
\providecommand \bibfnamefont [1]{#1}%
\providecommand \citenamefont [1]{#1}%
\providecommand \href@noop [0]{\@secondoftwo}%
\providecommand \href [0]{\begingroup \@sanitize@url \@href}%
\providecommand \@href[1]{\@@startlink{#1}\@@href}%
\providecommand \@@href[1]{\endgroup#1\@@endlink}%
\providecommand \@sanitize@url [0]{\catcode `\\12\catcode `\$12\catcode
  `\&12\catcode `\#12\catcode `\^12\catcode `\_12\catcode `\%12\relax}%
\providecommand \@@startlink[1]{}%
\providecommand \@@endlink[0]{}%
\providecommand \url  [0]{\begingroup\@sanitize@url \@url }%
\providecommand \@url [1]{\endgroup\@href {#1}{\urlprefix }}%
\providecommand \urlprefix  [0]{URL }%
\providecommand \Eprint [0]{\href }%
\providecommand \doibase [0]{http://dx.doi.org/}%
\providecommand \selectlanguage [0]{\@gobble}%
\providecommand \bibinfo  [0]{\@secondoftwo}%
\providecommand \bibfield  [0]{\@secondoftwo}%
\providecommand \translation [1]{[#1]}%
\providecommand \BibitemOpen [0]{}%
\providecommand \bibitemStop [0]{}%
\providecommand \bibitemNoStop [0]{.\EOS\space}%
\providecommand \EOS [0]{\spacefactor3000\relax}%
\providecommand \BibitemShut  [1]{\csname bibitem#1\endcsname}%
\let\auto@bib@innerbib\@empty
\bibitem [{\citenamefont {Hohenberg}\ and\ \citenamefont
  {Kohn}(1964)}]{hohenberg_inhomogeneous_1964}%
  \BibitemOpen
  \bibfield  {author} {\bibinfo {author} {\bibfnamefont {P.}~\bibnamefont
  {Hohenberg}}\ and\ \bibinfo {author} {\bibfnamefont {W.}~\bibnamefont
  {Kohn}},\ }\href {\doibase 10.1103/PhysRev.136.B864} {\bibfield  {journal}
  {\bibinfo  {journal} {Phys. Rev.}\ }\textbf {\bibinfo {volume} {136}},\
  \bibinfo {pages} {B864} (\bibinfo {year} {1964})}\BibitemShut {NoStop}%
\bibitem [{\citenamefont {Kohn}\ and\ \citenamefont
  {Sham}(1965{\natexlab{a}})}]{kohn_self-consistent_1965}%
  \BibitemOpen
  \bibfield  {author} {\bibinfo {author} {\bibfnamefont {W.}~\bibnamefont
  {Kohn}}\ and\ \bibinfo {author} {\bibfnamefont {L.~J.}\ \bibnamefont
  {Sham}},\ }\href {\doibase 10.1103/PhysRev.140.A1133} {\bibfield  {journal}
  {\bibinfo  {journal} {Phys. Rev.}\ }\textbf {\bibinfo {volume} {140}},\
  \bibinfo {pages} {A1133} (\bibinfo {year} {1965}{\natexlab{a}})}\BibitemShut
  {NoStop}%
\bibitem [{\citenamefont {Petkov}\ and\ \citenamefont
  {Stoitsov}(1991)}]{nuclear_dft}%
  \BibitemOpen
  \bibfield  {author} {\bibinfo {author} {\bibfnamefont {I.~Z.}\ \bibnamefont
  {Petkov}}\ and\ \bibinfo {author} {\bibfnamefont {M.~V.}\ \bibnamefont
  {Stoitsov}},\ }\href@noop {} {\emph {\bibinfo {title} {Nuclear Density
  Functional Theory}}}\ (\bibinfo  {publisher} {Oxford University Press},\
  \bibinfo {year} {1991})\BibitemShut {NoStop}%
\bibitem [{\citenamefont {Daily}\ \emph {et~al.}(2012)\citenamefont {Daily},
  \citenamefont {Rakshit},\ and\ \citenamefont
  {Blume}}]{PhysRevLett.109.030401}%
  \BibitemOpen
  \bibfield  {author} {\bibinfo {author} {\bibfnamefont {K.~M.}\ \bibnamefont
  {Daily}}, \bibinfo {author} {\bibfnamefont {D.}~\bibnamefont {Rakshit}}, \
  and\ \bibinfo {author} {\bibfnamefont {D.}~\bibnamefont {Blume}},\ }\href
  {\doibase 10.1103/PhysRevLett.109.030401} {\bibfield  {journal} {\bibinfo
  {journal} {Phys. Rev. Lett.}\ }\textbf {\bibinfo {volume} {109}},\ \bibinfo
  {pages} {030401} (\bibinfo {year} {2012})}\BibitemShut {NoStop}%
\bibitem [{\citenamefont {Thomas}(1927)}]{thomas_calculation_1927}%
  \BibitemOpen
  \bibfield  {author} {\bibinfo {author} {\bibfnamefont {L.~H.}\ \bibnamefont
  {Thomas}},\ }\href {\doibase 10.1017/S0305004100011683} {\bibfield  {journal}
  {\bibinfo  {journal} {Proc. Cambridege Philos. Soc.}\ }\textbf {\bibinfo
  {volume} {23}},\ \bibinfo {pages} {542} (\bibinfo {year} {1927})}\BibitemShut
  {NoStop}%
\bibitem [{\citenamefont {Fermi}(1928)}]{fermi_metodo_1927}%
  \BibitemOpen
  \bibfield  {author} {\bibinfo {author} {\bibfnamefont {E.}~\bibnamefont
  {Fermi}},\ }\href@noop {} {\bibfield  {journal} {\bibinfo  {journal} {Z.
  Phys.}\ }\textbf {\bibinfo {volume} {73}} (\bibinfo {year}
  {1928})}\BibitemShut {NoStop}%
\bibitem [{\citenamefont {Kirzhnits}(1957)}]{kirzhnits}%
  \BibitemOpen
  \bibfield  {author} {\bibinfo {author} {\bibfnamefont {D.~A.}\ \bibnamefont
  {Kirzhnits}},\ }\href@noop {} {\bibfield  {journal} {\bibinfo  {journal}
  {Sov. Phys. JETP}\ }\textbf {\bibinfo {volume} {5}},\ \bibinfo {pages} {64}
  (\bibinfo {year} {1957})}\BibitemShut {NoStop}%
\bibitem [{\citenamefont {Hodges}(1973)}]{hodges1973}%
  \BibitemOpen
  \bibfield  {author} {\bibinfo {author} {\bibfnamefont {C.~H.}\ \bibnamefont
  {Hodges}},\ }\href@noop {} {\bibfield  {journal} {\bibinfo  {journal}
  {Can. J. Phys.}\ }\textbf {\bibinfo {volume} {51}},\ \bibinfo
  {pages} {1428} (\bibinfo {year} {1973})}\BibitemShut {NoStop}%
\bibitem [{\citenamefont {Brack}\ \emph {et~al.}(1976)\citenamefont {Brack},
  \citenamefont {Jennings},\ and\ \citenamefont {Chu}}]{brack_extended_1976}%
  \BibitemOpen
  \bibfield  {author} {\bibinfo {author} {\bibfnamefont {M.}~\bibnamefont
  {Brack}}, \bibinfo {author} {\bibfnamefont {B.}~\bibnamefont {Jennings}}, \
  and\ \bibinfo {author} {\bibfnamefont {Y.}~\bibnamefont {Chu}},\ }\href
  {\doibase 10.1016/0370-2693(76)90519-0} {\bibfield  {journal} {\bibinfo
  {journal} {Phys. Lett. B}\ }\textbf {\bibinfo {volume} {65}},\ \bibinfo
  {pages} {1} (\bibinfo {year} {1976})}\BibitemShut {NoStop}%
\bibitem [{\citenamefont {Yang}(1986)}]{yang_gradient_1986}%
  \BibitemOpen
  \bibfield  {author} {\bibinfo {author} {\bibfnamefont {W.}~\bibnamefont
  {Yang}},\ }\href {\doibase 10.1103/PhysRevA.34.4575} {\bibfield  {journal}
  {\bibinfo  {journal} {Phys. Rev. A}\ }\textbf {\bibinfo {volume} {34}},\
  \bibinfo {pages} {4575} (\bibinfo {year} {1986})}\BibitemShut {NoStop}%
\bibitem [{\citenamefont {Kohn}\ and\ \citenamefont
  {Sham}(1965{\natexlab{b}})}]{kohn_quantum_1965}%
  \BibitemOpen
  \bibfield  {author} {\bibinfo {author} {\bibfnamefont {W.}~\bibnamefont
  {Kohn}}\ and\ \bibinfo {author} {\bibfnamefont {L.~J.}\ \bibnamefont
  {Sham}},\ }\href {\doibase 10.1103/PhysRev.137.A1697} {\bibfield  {journal}
  {\bibinfo  {journal} {Phys. Rev.}\ }\textbf {\bibinfo {volume} {137}},\
  \bibinfo {pages} {A1697} (\bibinfo {year} {1965}{\natexlab{b}})}\BibitemShut
  {NoStop}%
\bibitem [{\citenamefont {Elliott}\ \emph {et~al.}(2008)\citenamefont
  {Elliott}, \citenamefont {Lee}, \citenamefont {Cangi},\ and\ \citenamefont
  {Burke}}]{elliott_semiclassical_2008}%
  \BibitemOpen
  \bibfield  {author} {\bibinfo {author} {\bibfnamefont {P.}~\bibnamefont
  {Elliott}}, \bibinfo {author} {\bibfnamefont {D.}~\bibnamefont {Lee}},
  \bibinfo {author} {\bibfnamefont {A.}~\bibnamefont {Cangi}}, \ and\ \bibinfo
  {author} {\bibfnamefont {K.}~\bibnamefont {Burke}},\ }\href {\doibase
  10.1103/PhysRevLett.100.256406} {\bibfield  {journal} {\bibinfo  {journal}
  {Phys. Rev. Lett.}\ }\textbf {\bibinfo {volume} {100}},\ \bibinfo {pages}
  {256406} (\bibinfo {year} {2008})}\BibitemShut {NoStop}%
\bibitem [{\citenamefont {Kohn}\ and\ \citenamefont
  {Mattsson}(1998)}]{kohn_edge_1998}%
  \BibitemOpen
  \bibfield  {author} {\bibinfo {author} {\bibfnamefont {W.}~\bibnamefont
  {Kohn}}\ and\ \bibinfo {author} {\bibfnamefont {A.~E.}\ \bibnamefont
  {Mattsson}},\ }\href {\doibase 10.1103/PhysRevLett.81.3487} {\bibfield
  {journal} {\bibinfo  {journal} {Phys. Rev. Lett.}\ }\textbf {\bibinfo
  {volume} {81}},\ \bibinfo {pages} {3487} (\bibinfo {year}
  {1998})}\BibitemShut {NoStop}%
\bibitem [{\citenamefont {Engel}\ and\ \citenamefont
  {Dreizler}(2011)}]{engel_dreizler}%
  \BibitemOpen
  \bibfield  {author} {\bibinfo {author} {\bibfnamefont {E.}~\bibnamefont
  {Engel}}\ and\ \bibinfo {author} {\bibfnamefont {R.~M.}\ \bibnamefont
  {Dreizler}},\ }\href@noop {} {\emph {\bibinfo {title} {Density Functional
  Theory}}}\ (\bibinfo  {publisher} {Springer},\ \bibinfo {year}
  {2011})\BibitemShut {NoStop}%
\bibitem [{\citenamefont {Parr}\ and\ \citenamefont {Yang}(1989)}]{parr_yang}%
  \BibitemOpen
  \bibfield  {author} {\bibinfo {author} {\bibfnamefont {R.~G.}\ \bibnamefont
  {Parr}}\ and\ \bibinfo {author} {\bibfnamefont {W.}~\bibnamefont {Yang}},\
  }\href@noop {} {\emph {\bibinfo {title} {Density-Functional Theory of Atoms
  and Molecules}}}\ (\bibinfo  {publisher} {Oxford University Press},\ \bibinfo
  {year} {1989})\BibitemShut {NoStop}%
\bibitem [{\citenamefont {Weizs{\"a}cker}(1935)}]{weizsacker_1935}%
  \BibitemOpen
  \bibfield  {author} {\bibinfo {author} {\bibfnamefont {C.~F.~V.}\
  \bibnamefont {Weizs{\"a}cker}},\ }\href@noop {} {\bibfield  {journal}
  {\bibinfo  {journal} {Z. Phys.}\ }\textbf {\bibinfo {volume} {96}} (\bibinfo
  {year} {1935})}\BibitemShut {NoStop}%
\bibitem [{\citenamefont {Jennings}\ and\ \citenamefont
  {Bhaduri}(1975)}]{jennings_bhaduri}%
  \BibitemOpen
  \bibfield  {author} {\bibinfo {author} {\bibfnamefont {B.~K.}\ \bibnamefont
  {Jennings}}\ and\ \bibinfo {author} {\bibfnamefont {R.~K.}\ \bibnamefont
  {Bhaduri}},\ }\href@noop {} {\bibfield  {journal} {\bibinfo  {journal} {Nucl.
  Phys. A}\ }\textbf {\bibinfo {volume} {253}},\ \bibinfo {pages} {29}
  (\bibinfo {year} {1975})}\BibitemShut {NoStop}%
\bibitem [{\citenamefont {Vitos}\ \emph {et~al.}(2000)\citenamefont {Vitos},
  \citenamefont {Johansson}, \citenamefont {Koll\'ar},\ and\ \citenamefont
  {Skriver}}]{PhysRevA.61.052511}%
  \BibitemOpen
  \bibfield  {author} {\bibinfo {author} {\bibfnamefont {L.}~\bibnamefont
  {Vitos}}, \bibinfo {author} {\bibfnamefont {B.}~\bibnamefont {Johansson}},
  \bibinfo {author} {\bibfnamefont {J.}~\bibnamefont {Koll\'ar}}, \ and\
  \bibinfo {author} {\bibfnamefont {H.~L.}\ \bibnamefont {Skriver}},\ }\href
  {\doibase 10.1103/PhysRevA.61.052511} {\bibfield  {journal} {\bibinfo
  {journal} {Phys. Rev. A}\ }\textbf {\bibinfo {volume} {61}},\ \bibinfo
  {pages} {052511} (\bibinfo {year} {2000})}\BibitemShut {NoStop}%
\bibitem [{\citenamefont {Constantin}\ and\ \citenamefont
  {Ruzsinszky}(2009)}]{PhysRevB.79.115117}%
  \BibitemOpen
  \bibfield  {author} {\bibinfo {author} {\bibfnamefont {L.~A.}\ \bibnamefont
  {Constantin}}\ and\ \bibinfo {author} {\bibfnamefont {A.}~\bibnamefont
  {Ruzsinszky}},\ }\href {\doibase 10.1103/PhysRevB.79.115117} {\bibfield
  {journal} {\bibinfo  {journal} {Phys. Rev. B}\ }\textbf {\bibinfo {volume}
  {79}},\ \bibinfo {pages} {115117} (\bibinfo {year} {2009})}\BibitemShut
  {NoStop}%
\bibitem [{\citenamefont {Baltin}(1971)}]{baltin_1971}%
  \BibitemOpen
  \bibfield  {author} {\bibinfo {author} {\bibfnamefont {R.}~\bibnamefont
  {Baltin}},\ }\href@noop {} {\bibfield  {journal} {\bibinfo  {journal}
  {Phys. Lett. A}\ }\textbf {\bibinfo {volume} {37}},\ \bibinfo {pages}
  {67} (\bibinfo {year} {1971})}\BibitemShut {NoStop}%
\bibitem [{\citenamefont {Baltin}(1972)}]{baltin_1972}%
  \BibitemOpen
  \bibfield  {author} {\bibinfo {author} {\bibfnamefont {R.}~\bibnamefont
  {Baltin}},\ }\href@noop {} {\bibfield  {journal} {\bibinfo  {journal}
  {Naturforsch. A}\ }\textbf {\bibinfo {volume} {27}} (\bibinfo {year}
  {1972})}\BibitemShut {NoStop}%
\bibitem [{\citenamefont {Eguiluz}\ \emph {et~al.}(1992)\citenamefont
  {Eguiluz}, \citenamefont {Heinrichsmeier}, \citenamefont {Fleszar},\ and\
  \citenamefont {Hanke}}]{PhysRevLett.68.1359}%
  \BibitemOpen
  \bibfield  {author} {\bibinfo {author} {\bibfnamefont {A.~G.}\ \bibnamefont
  {Eguiluz}}, \bibinfo {author} {\bibfnamefont {M.}~\bibnamefont
  {Heinrichsmeier}}, \bibinfo {author} {\bibfnamefont {A.}~\bibnamefont
  {Fleszar}}, \ and\ \bibinfo {author} {\bibfnamefont {W.}~\bibnamefont
  {Hanke}},\ }\href {\doibase 10.1103/PhysRevLett.68.1359} {\bibfield
  {journal} {\bibinfo  {journal} {Phys. Rev. Lett.}\ }\textbf {\bibinfo
  {volume} {68}},\ \bibinfo {pages} {1359} (\bibinfo {year}
  {1992})}\BibitemShut {NoStop}%
\bibitem [{\citenamefont {Sahni}\ and\ \citenamefont
  {Solomatin}(1997)}]{sahni_1997}%
  \BibitemOpen
  \bibfield  {author} {\bibinfo {author} {\bibfnamefont {V.}~\bibnamefont
  {Sahni}}\ and\ \bibinfo {author} {\bibfnamefont {A.}~\bibnamefont
  {Solomatin}},\ }\href@noop {} {\bibfield  {journal} {\bibinfo  {journal}
  {Ann. Phys.}\ }\textbf {\bibinfo {volume} {259}},\ \bibinfo {pages}
  {97} (\bibinfo {year} {1997})}\BibitemShut {NoStop}%
\bibitem [{\citenamefont {Albright}(1977)}]{albright_1977}%
  \BibitemOpen
  \bibfield  {author} {\bibinfo {author} {\bibfnamefont {J.~R.}\ \bibnamefont
  {Albright}},\ }\href@noop {} {\bibfield  {journal} {\bibinfo  {journal} {J.
  Phys. A: Math. Gen.}\ }\textbf {\bibinfo {volume} {10}},\ \bibinfo {pages}
  {485} (\bibinfo {year} {1977})}\BibitemShut {NoStop}%
\bibitem [{\citenamefont {Lang}\ and\ \citenamefont
  {Kohn}(1970)}]{PhysRevB.1.4555}%
  \BibitemOpen
  \bibfield  {author} {\bibinfo {author} {\bibfnamefont {N.~D.}\ \bibnamefont
  {Lang}}\ and\ \bibinfo {author} {\bibfnamefont {W.}~\bibnamefont {Kohn}},\
  }\href {\doibase 10.1103/PhysRevB.1.4555} {\bibfield  {journal} {\bibinfo
  {journal} {Phys. Rev. B}\ }\textbf {\bibinfo {volume} {1}},\ \bibinfo {pages}
  {4555} (\bibinfo {year} {1970})}\BibitemShut {NoStop}%
\bibitem [{\citenamefont {Brack}\ and\ \citenamefont {van
  Zyl}(2001)}]{brack_simple_2001}%
  \BibitemOpen
  \bibfield  {author} {\bibinfo {author} {\bibfnamefont {M.}~\bibnamefont
  {Brack}}\ and\ \bibinfo {author} {\bibfnamefont {B.~P.}\ \bibnamefont {van
  Zyl}},\ }\href {\doibase 10.1103/PhysRevLett.86.1574} {\bibfield  {journal}
  {\bibinfo  {journal} {Phys. Rev. Lett.}\ }\textbf {\bibinfo {volume} {86}},\
  \bibinfo {pages} {1574} (\bibinfo {year} {2001})}\BibitemShut {NoStop}%
\bibitem [{\citenamefont {Brack}\ and\ \citenamefont
  {Murthy}(2003)}]{brack_harmonically_2003}%
  \BibitemOpen
  \bibfield  {author} {\bibinfo {author} {\bibfnamefont {M.}~\bibnamefont
  {Brack}}\ and\ \bibinfo {author} {\bibfnamefont {M.~V.~N.}\ \bibnamefont
  {Murthy}},\ }\href {\doibase 10.1088/0305-4470/36/4/318} {\bibfield
  {journal} {\bibinfo  {journal} {J. Phys. A: Math. Gen.}\ }\textbf {\bibinfo
  {volume} {36}},\ \bibinfo {pages} {1111} (\bibinfo {year}
  {2003})}\BibitemShut {NoStop}%
\bibitem [{\citenamefont {Armiento}\ and\ \citenamefont
  {Mattsson}(2002)}]{armiento_subsystem_2002}%
  \BibitemOpen
  \bibfield  {author} {\bibinfo {author} {\bibfnamefont {R.}~\bibnamefont
  {Armiento}}\ and\ \bibinfo {author} {\bibfnamefont {A.~E.}\ \bibnamefont
  {Mattsson}},\ }\href {\doibase 10.1103/PhysRevB.66.165117} {\bibfield
  {journal} {\bibinfo  {journal} {Phys. Rev. B}\ }\textbf {\bibinfo {volume}
  {66}},\ \bibinfo {pages} {165117} (\bibinfo {year} {2002})}\BibitemShut
  {NoStop}%
\bibitem [{\citenamefont {Hao}\ \emph {et~al.}(2010)\citenamefont {Hao},
  \citenamefont {Armiento},\ and\ \citenamefont
  {Mattsson}}]{PhysRevB.82.115103}%
  \BibitemOpen
  \bibfield  {author} {\bibinfo {author} {\bibfnamefont {F.}~\bibnamefont
  {Hao}}, \bibinfo {author} {\bibfnamefont {R.}~\bibnamefont {Armiento}}, \
  and\ \bibinfo {author} {\bibfnamefont {A.~E.}\ \bibnamefont {Mattsson}},\
  }\href {\doibase 10.1103/PhysRevB.82.115103} {\bibfield  {journal} {\bibinfo
  {journal} {Phys. Rev. B}\ }\textbf {\bibinfo {volume} {82}},\ \bibinfo
  {pages} {115103} (\bibinfo {year} {2010})}\BibitemShut {NoStop}%
\bibitem [{\citenamefont {Thorpe}\ and\ \citenamefont
  {Thouless}(1970)}]{thorpe_thouless}%
  \BibitemOpen
  \bibfield  {author} {\bibinfo {author} {\bibfnamefont {M.~A.}\ \bibnamefont
  {Thorpe}}\ and\ \bibinfo {author} {\bibfnamefont {D.~J.}\ \bibnamefont
  {Thouless}},\ }\href@noop {} {\bibfield  {journal} {\bibinfo  {journal}
  {Nucl. Phys. A}\ }\textbf {\bibinfo {volume} {156}},\ \bibinfo {pages} {225}
  (\bibinfo {year} {1970})}\BibitemShut {NoStop}%
\bibitem [{\citenamefont {Pearson}\ and\ \citenamefont
  {Gordon}(1985)}]{pearson_symptotoc}%
  \BibitemOpen
  \bibfield  {author} {\bibinfo {author} {\bibfnamefont {E.~W.}\ \bibnamefont
  {Pearson}}\ and\ \bibinfo {author} {\bibfnamefont {R.~G.}\ \bibnamefont
  {Gordon}},\ }\href@noop {} {\bibfield  {journal} {\bibinfo  {journal} {J.
  Chem. Phys}\ }\textbf {\bibinfo {volume} {82}} (\bibinfo {year}
  {1985})}\BibitemShut {NoStop}%
\bibitem [{\citenamefont {Kleinman}\ and\ \citenamefont
  {Lee}(1988)}]{PhysRevB.37.4634}%
  \BibitemOpen
  \bibfield  {author} {\bibinfo {author} {\bibfnamefont {L.}~\bibnamefont
  {Kleinman}}\ and\ \bibinfo {author} {\bibfnamefont {S.}~\bibnamefont {Lee}},\
  }\href {\doibase 10.1103/PhysRevB.37.4634} {\bibfield  {journal} {\bibinfo
  {journal} {Phys. Rev. B}\ }\textbf {\bibinfo {volume} {37}},\ \bibinfo
  {pages} {4634} (\bibinfo {year} {1988})}\BibitemShut {NoStop}%
\bibitem [{\citenamefont {Armiento}\ and\ \citenamefont
  {Mattsson}(2003)}]{PhysRevB.68.245120}%
  \BibitemOpen
  \bibfield  {author} {\bibinfo {author} {\bibfnamefont {R.}~\bibnamefont
  {Armiento}}\ and\ \bibinfo {author} {\bibfnamefont {A.~E.}\ \bibnamefont
  {Mattsson}},\ }\href {\doibase 10.1103/PhysRevB.68.245120} {\bibfield
  {journal} {\bibinfo  {journal} {Phys. Rev. B}\ }\textbf {\bibinfo {volume}
  {68}},\ \bibinfo {pages} {245120} (\bibinfo {year} {2003})}\BibitemShut
  {NoStop}%
\bibitem [{\citenamefont {Uns{\"o}ld}(1927)}]{unsold_theorem}%
  \BibitemOpen
  \bibfield  {author} {\bibinfo {author} {\bibfnamefont {A.}~\bibnamefont
  {Uns{\"o}ld}},\ }\href@noop {} {\bibfield  {journal} {\bibinfo  {journal}
  {Ann. Phys. (Berlin)}\ }\textbf {\bibinfo {volume} {82}} (\bibinfo {year}
  {1927})}\BibitemShut {NoStop}%
\end{thebibliography}

%
%

%

\end{document}